\documentclass[11pt]{article}
\pdfoutput=1
\usepackage{empheq}

\usepackage{amsmath, amsfonts, amssymb}
\usepackage{comment}
\usepackage{graphicx}
\usepackage{psfrag}
\usepackage{enumerate}
\usepackage{soul}
\usepackage{subfig}
\usepackage{cite}
\usepackage{a4wide}
\usepackage{tikz}
\usepackage{slashed}

\usepackage{float}

\usepackage{color}
\definecolor{dark-gray}{gray}{0.20}
\definecolor{gray}{gray}{0.30}
\definecolor{light-gray}{gray}{0.80}
\definecolor{dark-red}{rgb}{0.7,0,0}
\definecolor{dark-green}{rgb}{0.1,0.4,0}
\definecolor{dark-blue}{rgb}{0.3,0.3,0.7}
\definecolor{light-blue}{rgb}{0.8,0.8,1}

\usepackage{blindtext}
\usepackage{tcolorbox}
\usepackage{xcolor}
\usepackage{framed}
\colorlet{shadecolor}{blue!10}

\usepackage{hyperref}
\usepackage{setspace}
\hypersetup{
    colorlinks=true,
    linkcolor=dark-blue,
    citecolor=dark-red,
    urlcolor=dark-green,
    linktoc=page
}

\numberwithin{equation}{section}

\def\be{\begin{equation}}
\def\ee{\end{equation}}
\def\bea{\begin{eqnarray}}
\def\eea{\end{eqnarray}}

\newcommand{\vol}{\mathrm{vol}}

\def \dd  {{\rm d}}

\usepackage{letltxmacro}
\makeatletter
\let\oldr@@t\r@@t
\def\r@@t#1#2{%
\setbox0=\hbox{$\oldr@@t#1{#2\,}$}\dimen0=\ht0
\advance\dimen0-0.2\ht0
\setbox2=\hbox{\vrule height\ht0 depth -\dimen0}%
{\box0\lower0.4pt\box2}}
\LetLtxMacro{\oldsqrt}{\sqrt}
\renewcommand*{\sqrt}[2][\ ]{\oldsqrt[#1]{#2}}
\makeatother

\usepackage{eso-pic}
\newcommand{\reportnum}[2]{
  \AddToShipoutPictureBG*{%
    \AtPageUpperLeft{%
      \hspace{0.75\paperwidth}%
      \raisebox{#1\baselineskip}{%
        \makebox[0pt][l]{\textnormal{#2}}
  }}}%
}

\begin{document}
\reportnum{-5}{MPP-2025-62}

\begin{center}
    {\LARGE\bf Black Holes, Moduli Stabilisation\\\vspace{0.3cm} and the Swampland}\\

    \vspace{1.5cm}
    {\large M. Delgado$^{1,2}$, S. Reymond$^3$, T. Van Riet$^3$}\\
    \vspace{1cm}

    \emph{$^1$Max-Planck-Institut f\"ur Physik (Werner-Heisenberg-Institut)\\
Boltzmannstr. 8, 85748 Garching, Germany}\\\vspace{0.15cm}

\emph{$^2$Jefferson Physical Laboratory, Harvard University\\ 17 Oxford St, Cambridge, MA 02138, United States of America}
    \\ \vspace{0.15cm}
    {$^3$\emph{
        Instituut voor Theoretische Fysica, K.U. Leuven,\\
        Celestijnenlaan 200D B-3001 Leuven, Belgium
    }}
    
    \vspace{2cm}
    
    {\bf Abstract}
\end{center}

{\small In theories with moduli, extremal black holes exhibit a striking feature: for generic initial conditions, the distance travelled by the scalars from infinity to the horizon can grow with the size of the black hole, allowing larger black holes to probe more of the theory’s UV ingredients. This apparent tension with naive EFT expectations is related to the absence of cosmological moduli stabilisation. Indeed, in string vacua with parametrically heavy stabilised scalars—dubbed \emph{rigid} compactifications—one recovers the EFT intuition that only small black holes probe the UV. Crucially, “small” here means small compared to the Compton wavelength $1/m$ of the stabilised moduli.
If moduli stabilisation is not rigid, as is the case in all controlled string vacua, then even astrophysical black holes are small in this sense. Hence, although the distances travelled typically decrease with black hole size, they can still span many Planck units even for black holes that are enormous in Planck units. We first demonstrate this explicitly in a few toy models. We then construct near-horizon solutions in a top-down, scale-separated type IIA compactification with stabilised moduli, where sizeable field variations occur for large black holes because the moduli are not parametrically heavy. These results lead us to speculate that rigid compactifications belong to the Swampland if one requires UV physics near the horizons of large black holes.}

\setcounter{tocdepth}{2}

\newpage
\tableofcontents
\newpage

\section{Introduction}

Our understanding of (quantum) gravity has advanced significantly through Gedanken experiments involving black hole spacetimes, where UV/IR mixing is exploited to uncover low-energy traces of quantum gravitational phenomena. Whether we can go beyond Gedanken experiments and use the actual direct detection of black holes established in recent years to shed light on observational aspects of quantum gravity remains to be seen.

At first sight, it might seem surprising that black holes have the ability to teach us anything about quantum gravity if we study their horizon. After all, black holes at horizon scales do not have to be dense objects or feature strong curvature. The larger a black hole, the weaker the curvature of spacetime near its horizon, and the smaller its energy density. For black holes in the center of our galaxy the density is not much more than that of water, for instance. Or even more dramatically; when all matter in the observable universe would be pressed into a black hole, the black hole radius would still be on the order of the radius of the current universe. So standard \emph{Effective Field Theory} (EFT) reasoning suggests that we should be in a regime where quantum gravity is negligible, as its effects are expected to be Planck-suppressed and so accessing the quantum gravity regime directly would require curvature scales or energy densities to reach Planckian levels.

On the other hand, there is strong theoretical evidence that black holes are intrinsically quantum in nature. A prominent insight obtained from pure thought is the black hole information paradox; see \cite{Mathur:2009hf, Almheiri:2020cfm,Raju:2020smc} for some modern reviews. The starting point is the assumption (the \emph{central dogma}) that from the outside, a black hole should look like a quantum system made out of microscopic degrees of freedom that evolve unitarily under time evolution. The sheer fact that the horizon area is an entropy counting microstates implies general relativity fails to describe what happens, even far away from the singularity. Interestingly, string theory has provided us with examples where microstate counting can be verified, including detailed matching of numerical coefficients \cite{Strominger:1996sh}. This shows black holes are theoretical testing grounds for the consistency of gravity theories \cite{Sen:2012dw}. 

Furthermore, the various proposals for the resolution of the information paradox inform us that quantum gravity near a black hole horizon must be sufficiently non-local in order for black hole evaporation to be unitary. How this non-locality is implemented depends on the would-be microscopic realization of each proposal, which is not always understood. When the microscopic realization in string theory is understood, it seems the non-locality is due to UV structures that arise at the  horizon \cite{Mathur:2005zp, Skenderis:2008qn}.  An extreme example of this, in specific String Theory set-ups \emph{without} moduli stabilisation, are fuzzballs.
 For a fuzzball the extra dimensions become so floppy near the horizon region that the entire 10-dimensional picture becomes relevant (the KK tower becomes light) and removes both horizon and singularity. Indeed, as explained in Appendix  \ref{App:FUZZ} fuzzballs generally explore \emph{infinite distance in moduli space} (see also \cite{Li:2021utg}).

In this sense black holes can mix the UV and the IR and challenge our notion of EFT. This explains why they have been so pivotal in developing the Swampland program \cite{Vafa:2005ui, Palti:2019pca}. This program aims at classifying EFTs into two classes; those that can be UV completed (EFTs in the landscape) and the rest (EFTs in the Swampland). This classification contains information about quantum gravity beyond what EFT can teach us. A famous example of a Swampland principle that was first uncovered using black hole thought experiments is the absence of global symmetries in quantum gravity \cite{Palti:2019pca}. 

Most insights into black holes from string theory were obtained in special circumstances that have little to do with (black holes observed in) our own universe. First, the black holes studied in detail tend to be supersymmetric, or at least extremal. This feature is used for the needed computational control; not only for constructing the solutions, but also for the counting of its microstates. Second, these black holes live inside universes that do not look at all like our own; there are many exactly massless fields (moduli) that couple to all the matter. We do not observe such moduli in our universe, and so we should study string theoretic universes that originate from compactifications without moduli. But that is not enough. We also want the KK modes to be much heavier than the AdS scale, a feature called \emph{scale separation}, since otherwise an observer would not consider the universe to be four-dimensional. What is needed are therefore vacua where moduli obtain a mass significantly larger than the cosmological constant scale. We will refer to them as \emph{rigid} compactifications:\\

\underline{Definition:} \emph{A rigid compactification has a large separation between Kaluza-Klein (KK) and (A)dS scale, $L_{KK}/L_{(A)dS} \ll 1$, and all scalar field masses $m$ are  large in (A)dS units  $mL_{(A)dS} \gg 1$.\footnote{An equivalent statement is that all KK masses are large in AdS units, even that of the scalar zero modes.}}\\

In a rigid compactification, one can define an EFT below the KK scale where the moduli are stabilized at their vev. Clearly, the wish list to find vacua that look like our own universe is much longer; moduli stabilisation and separation is a very basic requirement for a realistic universe and yet surprisingly hard to achieve \cite{Coudarchet:2023mfs, VanRiet:2023pnx} to the extent it is still being debated whether we have sufficient evidence for a landscape of scale-separated vacua. The condition of rigidity is perhaps not needed but then one needs a mechanism to have suppressed couplings of the scalars to all other matter. The difficulty in avoiding light scalars is encapsulated in the AdS moduli conjecture \cite{Gautason:2018gln}.

In this paper we wish to relate how black holes behave in semi-realistic vacua to the debate on the consistency of rigid compactifications. As black holes pave the way for most Swampland principles, it is possible that the study of black holes in would-be vacua sheds light on the consistency of these vacua.  This goal is ambitious, especially given the lack of earlier studies in this direction. To our knowledge, only two papers have studied black hole solutions with cosmological moduli stabilisation in stringy EFTs before: Green, Silverstein and Starr in \cite{Green:2006nv} and Danielsson, Johansson and Larfors in \cite{Danielsson:2006jg}. So, even the very existence of black hole solutions in would-be scale-separated vacua is not properly understood. Hence, one of the goals of this paper is to present a simple near-horizon solution in a IIA flux vacuum that is claimed to achieve scale separation and full moduli stabilisation \cite{DeWolfe:2005uu} (this vacuum was also simultaneously found in \cite{Camara:2005dc}).

\begin{figure}
    \centering
    \includegraphics[width=0.8\linewidth]{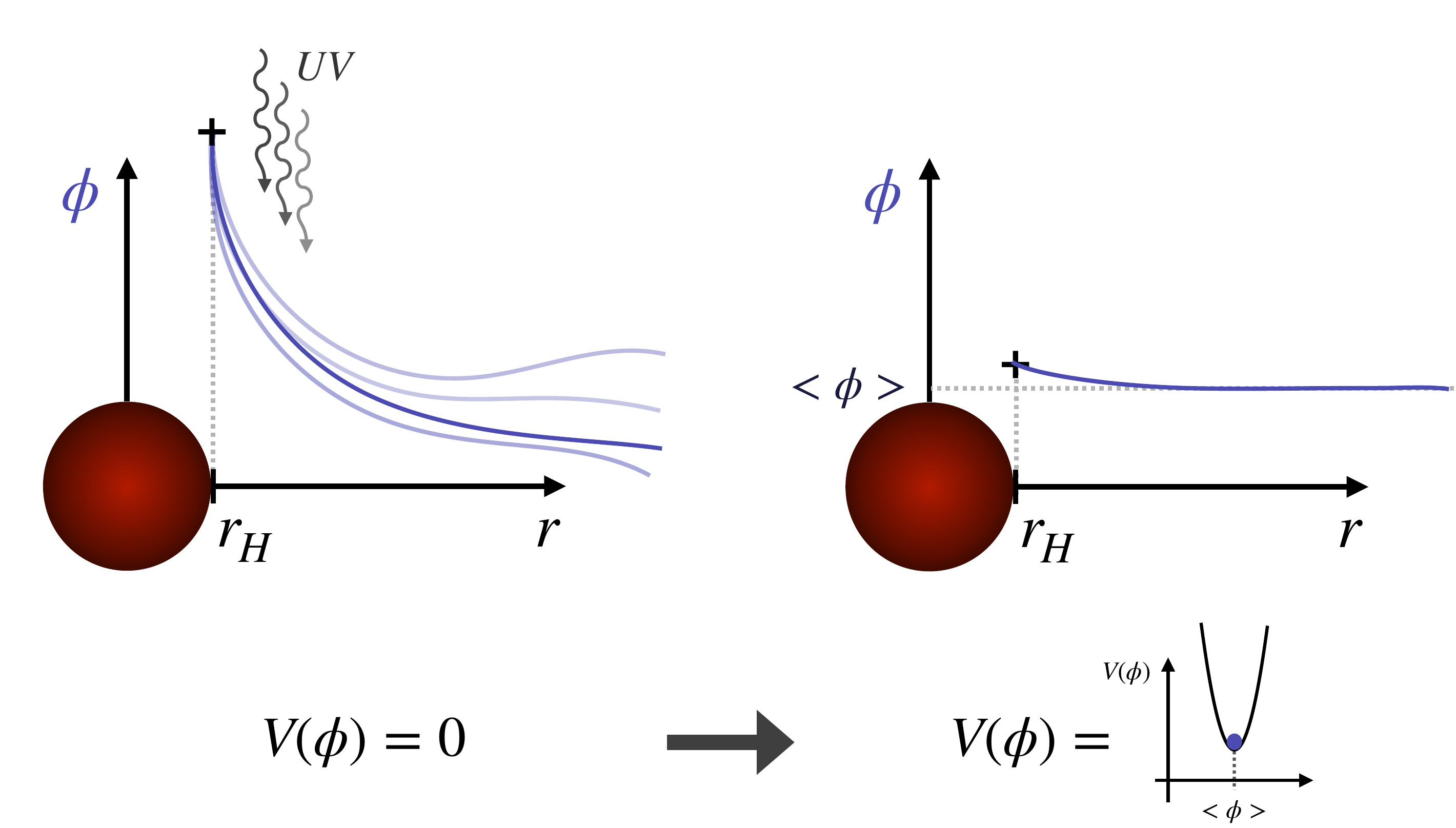}\vspace{1cm}
    \caption{\emph{On the left: the attractor mechanism in absence of a scalar potential. The black hole drags the scalar to a specific value, dictated by the charges. 
    On the right: the same black hole but in the presence of a steep stabilizing potential for the scalar. The steeper the potential, the more rigid the compactification and the harder it is for the black hole to drag the scalar off of its cosmological minimum. In turn, UV physics will not appear at horizon scales.}  }
    \label{fig:illustr}
\end{figure}

Our main focus will be on the behavior of scalar fields near extremal black holes.  In generic Minkowski vacua without moduli stabilisation the horizon is an attractor point for the moduli \cite{Ferrara:1996dd}: independently of their value at spatial infinity they flow to a fixed point at the horizon determined solely by the charges of the black hole. The distance $\Delta$ traveled in moduli space from the moduli vev at spatial infinity towards the horizon can be chosen such that they are made large as one makes the BH mass $M$ large, roughly as (see e.g. \cite{Bonnefoy:2019nzv,Delgado:2022dkz}):
\begin{equation}
\Delta \sim \log(M/M_{pl})\,.    
\end{equation}
Hence, large black hole can probe large distances in moduli space. Through the distance conjecture \cite{Ooguri:2006in} this means that large black holes probe more UV properties of the theory since a tower of UV states becomes light exponentially in $\Delta$. Indeed, one of the most important aspects of the Swampland program is that \textbf{quantum gravitational effects are not suppressed by the Planck scale on its own}: instead, UV scales depend on the value of the moduli and become light when these moduli become large. This behavior is an example of how naive EFT properties are being challenged since large black holes have lower curvature and are not expected to probe the UV of the theory near its horizon. For this reason, large black holes have been used in the Swampland literature to probe the UV and study the link between the tower states and the microstates of the black hole \cite{Bonnefoy:2019nzv,Cribiori:2022cho,Cribiori:2022nke,Cribiori:2023ffn,Basile:2023blg,Basile:2024dqq,Herraez:2024kux,Calderon-Infante:2025pls}. This case is depicted on the left of Figure \ref{fig:illustr}.

On the other hand, in the presence of a cosmological scalar potential, scalars are stabilized at some UV scale $\Lambda_{UV}$. If $\Lambda_{UV}$ is very high, we do not expect black holes to be able to significantly displace the scalars away from their cosmological minimum unless the black holes become so small that its radius becomes of the order of $\Lambda_{UV}^{-1}$. Indeed, whilst large black holes without moduli stabilization \textbf{can} drag moduli to large distances, once the scalars are massive, this is no longer an option. In fact, at least in directions in which the potential grows away from the minimum, the larger the black hole, the smaller the distance traveled by the massive scalars to the horizon. Therefore, in these scenarios large black holes cannot probe the UV of the theory and hence we expect the absence of sizable quantum gravitational effects near the horizon, as depicted on the right in Figure \ref{fig:illustr}. This fact would have drastic implications for the proposed resolutions of the information paradox that involve UV structures at the horizon. For instance, as we discuss below in Appendix \ref{App:FUZZ}, the fuzzball proposal is in clear tension with having a rigid vacuum. Therefore, if one needs UV structures at the horizon to solve the information paradox, then this would be evidence that rigid vacua are in the Swampland, consistent with the conjecture of \cite{Gautason:2018gln}. At the phenomenological level, this would imply that new physics should become manifest (at least) near the horizon of large black holes, which could be in the form of a large yet-unnoticed dark sector.

Since it is central to the paper, and since the terminology is used differently in the literature, let us be explicit about what we mean by ``small'' and ``large'' black holes. Throughout this paper a black hole of horizon radius $R$ is called \emph{small} if $mR\ll 1$ and \emph{large} if $mR\gg1$, where $m$ is the mass of the stabilised modulus in the vacuum. We stress two things:
\begin{itemize}
    \item Smallness is measured with respect to $m$, \emph{not} with respect to the Planck scale. A black hole can be astronomically large in Planck units and still be small in our sense. For instance, if the moduli masses are of the order of the Hubble scale, $m\sim H_0$, then Sagittarius A$^*$, with $R\sim 10^{10}\,$m, has $mR\sim 10^{-16}$ and is a small black hole.
    \item Our small black holes have a macroscopic horizon and are large in Planck units. They should not be confused with the singular, vanish-horizon-area solutions that are also sometimes called ``small black holes'' in the literature (see e.g. \cite{Dabholkar:2012zz}). We never consider the latter.
\end{itemize}
With this terminology, the picture that emerges is the following. Without a potential, black holes of \emph{any} size can drag the moduli arbitrarily far (the distance grows with the charges). With a stabilising potential, only small black holes, $mR\lesssim 1$, can do so, and the distance is finite for any black hole of finite size. But since in all known string compactifications the moduli are \emph{not} parametrically heavy compared to the (A)dS scale (see the AdS moduli conjecture \cite{Gautason:2018gln}), the class of ``small'' black holes may include all astrophysical ones. They can then still drag the moduli over many Planck units, although never to infinite distance. This is what we find in the two toy models where the scalar mass is tied to the AdS scale (Models 2 and 3 below): black holes smaller than the AdS radius drag the scalar over many Planck units, larger ones displace it by an $\mathcal{O}(1)$ amount. Only in rigid compactifications does the class of small black holes shrink to (sub)Planckian sizes, in which case the naive EFT expectation is recovered. There is one exception, in Model 3, whose potential vanishes at weak coupling as string potentials do: for suitable charges, black holes much larger than the AdS radius drag the modulus arbitrarily far along this flat direction, with a distance growing logarithmically with their size. In the top-down model of section~\ref{sec:topdown}, we confront the toy models to explicit near-horizon geometries in a scale-separated DGKT vacuum. We only find horizons larger than the AdS radius. They show both large-black-hole behaviours: $\mathcal{O}(1)$ displacements, or a run-away to weak coupling and large volume as in Model 3. Within our truncation all of them are perturbatively unstable.

The rest of this paper is organised as follows. In section \ref{sec:flows} we review the attractor mechanism without a potential and present the formalism for scalar field flows of charged black holes with cosmological moduli stabilisation. In section \ref{sect:bottom_up_EMD_theory} we study three bottom-up toy models: a mass term around a Minkowski vacuum, and two potentials with AdS vacua, one of them with string-like asymptotics. In section \ref{sec:topdown} we construct near-horizon solutions in a IIA scale-separated flux vacuum on $\mathbb{T}^6/\mathbb{Z}_4$ and analyse their stability. We speculate on the implications for the Swampland program and the resolution of the information paradox in section \ref{sec:discussion}. The appendices contain further technical results used in the main text.

\section{Black Hole Flow Equations}\label{sec:flows}
In this section we explore spherical black hole solutions with charges. First, we recall (in words)  the attractor mechanism for charged extremal black holes without cosmological moduli stabilisation and how large black holes can be used to probe UV physics in this setup. Then, in section \ref{sect:BH_flow_scalar_potential}, we present the black hole flow equations in the presence of a  scalar potential. This sets the stage for section \ref{sect:bottom_up_EMD_theory}, where we apply these equations to a few toy models: Einstein-Maxwell-dilaton theory with a $\phi^2$ mass term for the dilaton (Minkowski vacuum), and with two potentials that have an AdS vacuum, one of which has string-like asymptotics. In section \ref{sec:topdown} we discuss top-down examples.

\subsection{Scalar Field Distances for Charged Black Holes (and No Potential)}

The attractor mechanism for black holes (see for instance \cite{DallAgata:2011zkh}) occurs  in the context of effective field theories with massless scalar fields that couple to Abelian gauge fields. Let us consider the simplest case of Einstein Maxwell theory with a single scalar:
\begin{equation}\label{eq:schemat}
    S=\frac{1}{2\kappa_d^2}\int dx^d  \sqrt{-g}  \left(R-\tfrac{1}{2} (\partial\phi)^2 -\tfrac{1}{4} e^{a\phi} |F|^2\right)  \,. 
\end{equation}
This is a schematic example of the type of EFTs that can be easily obtained from string theory: some strictly massless scalar fields, represented by $\phi$ and some U(1) gauge fields, whose kinetic term is represented by $|F|^2$. Importantly, the gauge fields couple to the scalars through the gauge kinetic function $e^{a \phi}$ for $a$ some $\mathcal{O}(1)$ parameter which is model dependent. This is what implements the attractor mechanism: any black hole solution that is charged under these U(1) symmetries will source an effective potential for the scalars $\phi$. It can be shown that the charges of the black hole entirely determine the value of $\phi$ at the horizon, independently of its value infinitely far away from the black hole. This profile for the scalar is usually called \emph{secondary} scalar hair. We refer the reader who is not familiar with the attractor mechanism to the beginning of section \ref{sect:BH_flow_scalar_potential}, where we describe it in detail.

Such a set up has been used at various instances in the Swampland literature to explore whether or not these black holes can probe UV physics at large distances in moduli space. In general, two regimes were considered; in the first, singular (so-called \textit{small}) black holes were considered.\footnote{This is the \emph{only} place where ``small black hole'' refers to a singular, zero-area solution. In the rest of the paper small and large are always measured relative to the inverse mass $1/m$ of the stabilised moduli, and all black holes we consider have a regular, macroscopic horizon.} These solutions are curvature singularities, where the moduli naively blow up and they can be understood as black holes that do not carry enough charges to generate a macroscopic event horizon. At their singular core, the EFT breaks down completely, and UV effects become manifest. In fact, it was argued that these effects could come in to smooth out the singularity; this was done by studying how stringy curvature corrections affect the singular solution (see e.g. \cite{Dabholkar:2012zz} for a review and \cite{Calderon-Infante:2023uhz} in a Swampland context). It should not come as a surprise that such singular objects would probe UV physics and the competition of this effect with a scalar potential has been studied in \cite{Angius:2023xtu}.

Perhaps more surprising is the other regime in which one can probe UV effects in stringy EFTs. In 4d and 5d supergravity theories obtained from compactifications of string and M-theory, BPS attractors have been used to drag the moduli to the boundary of moduli space on the horizon of very large black holes. Indeed, the attracted value of the moduli is invariant under a homogeneous rescaling of all the charges, but if one picks a ratio of the charges to be very large $Q_i/Q_j\gg1$, one can make the attracted value of the moduli very large. This usually comes at the price of taking some of the charges to be near infinite $Q_i\gg1$ (since one cannot make charges arbitrarily small due to Dirac quantization $|Q_j|\geq1$), which has the effect of making the black hole itself very large in Planck units. For instance, such black holes were studied in \cite{Bonnefoy:2019nzv}, where it was argued that a tower of light states is expected to emerge in the limit of large black hole entropy. Moreover, in \cite{Delgado:2022dkz}, BPS black holes were argued to encode UV data about the underlying compactification in their thermodynamic properties. From the perspective of an effective field theorist, it is surprising that such large and smooth black holes should probe UV effects. Indeed, as the horizon gets large, one expects curvature corrections to become suppressed, and the black hole to be well within the regime of a classical description. The resolution of this apparent puzzle is that although these black holes are very large in Planck units, they are still comparable to the UV cut-off scales of the EFT. Indeed, one of the main results of the Swampland program has been the realization that stringy EFTs break down at scales exponentially smaller than the Planck scale at infinite distance in moduli space. This is encapsulated in the Distance conjecture \cite{Ooguri:2006in} that states that an infinite tower of states becomes exponentially light in such limits. The emergent string conjecture \cite{Lee:2019wij} further argues that this tower of states is always either the tower of Kaluza-Klein modes associated to a decompactification to a higher dimensional theory or to the tower of oscillation modes of a weakly coupled (dual) string. This infinite number of modes becoming massless signals the break down of the theory. This is captured by the lowering of the \emph{species scale} (see \cite{Dvali:2007hz,Dvali:2007wp,Dvali:2008ec,Dvali:2009ks,Dvali:2010vm,Dvali:2012uq} for early references and \cite{Castellano:2022bvr , Castellano:2023aum} for obtaining the species scale from counting tower states in the infinite distance limit), which can be thought of as the energy scale at which quantum gravitational effects become relevant and where higher-curvature corrections appear (see \cite{vandeHeisteeg:2022btw,vandeHeisteeg:2023dlw} and in relation to black holes \cite{Cribiori:2022nke ,Calderon-Infante:2023uhz  ,Cribiori:2023ffn, Herraez:2024kux ,Basile:2024dqq,Bedroya:2024ubj,Calderon-Infante:2025ldq ,Castellano:2025ljk,Basile:2023blg }). As one goes to infinite distance in moduli space in a d-dimensional EFT of string theory, the species scale becomes exponentially light in d-dimensional Planck units: 
\begin{equation}
    \frac{\Lambda_s}{M_{pl,d}} \sim e^{-\alpha \Delta}\;\;\; \text{as}\,\;\, \Delta\to \infty\,,
\end{equation}
where $\Delta$ is the geodesic distance in moduli space and $\alpha$ is a positive $\mathcal{O}(1)$ parameter that depends on the specific infinite distance limit that is probed. Therefore, as one uses very large black holes to probe infinite distances in moduli space at its horizon, the length scale at which quantum gravity becomes manifest becomes large at the horizon as well. If the black hole we consider is large in Planck units but $\mathcal{O}(1)$ in species scale units, this would mean that the black hole cannot be treated classically and has UV physics playing a role at its horizon. It becomes clear that one should always compare the sizes of these black holes to the species scale instead of the Planck scale. 

In fact it was recently argued in \cite{Calderon-Infante:2025pls}, that the smallest such large black holes are often species scale sized. More precisely, when one sets all of the extra charges to $\mathcal{O}(1)$ numbers (of course, not the ones $Q_i\gg1$ that are responsible for dragging the moduli to the boundary of moduli space), one finds that the resulting black hole is comparable to the species scale (or another UV scale, like the KK scale in some cases). Therefore, for general $\mathcal{O}(1)$ values of these extra charges, the resulting black hole is literally KK scale or species-scale-sized. Of course, if one takes these extra charges to infinity as well, the black hole can be larger than these scales, but the horizon still takes the moduli to infinite distance in moduli space, signaling that UV scales are becoming light. 

Summing up, since UV scales become light at infinite distance in moduli space, and since the attracted value of the scalars can be made very large with large charges, UV physics is at play at the horizon of large black holes. In the next section, we will see how this picture changes in presence of a scalar potential for the moduli.

\subsection{Black Hole Flow Equations with a Scalar Potential}
\label{sect:BH_flow_scalar_potential}

We now turn to describing how the attractor mechanism is modified in the presence of a scalar potential. 

Consider the following generic action of gravity coupled to scalars and abelian vectors
\begin{equation}
S = \frac{1}{2\kappa_4^2}\int d^4 x \sqrt{-g}\left( \mathcal{R} - \frac{1}{2}G_{ij}(\phi)\partial\phi^i \partial\phi^j - \frac{1}{4}A_{IJ}F^I_{\mu\nu} F^{J\mu\nu} - \frac{1}{8}B_{IJ}F^I_{\mu\nu} F^J_{\rho\sigma}\epsilon^{\mu\nu\rho\sigma} -  V(\phi)\right). \label{eq:generic_gravity_action}
\end{equation}
Here $G_{ij}$ is the metric on the scalar manifold, $F^I=\dd A^I$ with $A^I$ an abelian vector, and the matrices $A, B$ are functions of the scalars. Static and spherical black holes  obey the following ansatz:
\begin{align}
   & ds^2 = -e^{2U(z)} \dd t^2 + e^{-2U(z)}\left[e^{4A(z)}\dd z^2 + e^{2A(z)}\dd \Omega^2\right]\,, \label{eq:metric_ansatz_flow_eq}\\
   & F^I = P^I d\Omega_2 + e^{2U}(A^{-1})^{IM}\left(-Q_M + B_{MN}P^N\right) \dd t \wedge \dd z \,,\\
   & \phi^I = \phi^I(z) \,.
\end{align}
The Bianchi identity and equations of motion for the vectors tell us that $P^I$ and $Q_I$ are constants, corresponding to magnetic and electric charges. When we restrict ourselves to static and spherical black hole solutions, we can dimensionally reduce the above to an effective one-dimensional action which captures all the equations of motion:
\begin{align}
    S_{\text{eff}} &\simeq \frac{1}{2\kappa_4^2}\int \dd z \, \left(2\dot{U}^2 - 2\dot{A}^2 - 2e^{2A} + \frac{1}{2}G_{ij}\dot{\phi}^i\dot{\phi}^j + e^{2U} V_{BH}(\phi) + e^{-2U+4A}V(\phi) \right) \,, \label{eq:S_eff_generic_fields}
\end{align}
together with the Hamiltonian constraint
\begin{align}
    2\dot{U}^2 - 2\dot{A}^2 + \frac{1}{2}G_{ij}\dot{\phi}^i\dot{\phi}^j = - 2e^{2A} + e^{2U} V_{BH}(\phi) + e^{-2U+4A}V(\phi)\, . \label{eq:Hamiltonian_constraint_generic_fields}
\end{align}
Here $V_{BH}$ is the black hole potential sourced by the charges $Q_I, P^I$ and is given by
\begin{equation}
    V_{BH}(\phi) = \frac{1}{2}A^{IJ} Q_I  Q_J  - A^{IJ} Q_I P_J + \frac{1}{2}A_{IJ} P^I P^J + \frac{1}{2}A^{MN} P_M P_N\,. \label{eq:BH_potential_generic}
\end{equation}
The notation used is that $A$ with indices upstairs denotes the inverse of $A$ and $P$ with indices downstairs is defined as $P_I= B_{IJ}P^J$\footnote{Note that $A_{IJ} P^I P^J \neq A^{IJ} P_I P_J$ since the indices of $P$ are lowered using the $B$ matrix and $A^{IJ}$ denotes components of the inverse of $A$.}. The equations of motion of \eqref{eq:S_eff_generic_fields} together with the constraint \eqref{eq:Hamiltonian_constraint_generic_fields} define the flow equations. One can already see the competition between the effect of the black hole and that of the scalar potential on the scalars in \eqref{eq:S_eff_generic_fields}. 

\paragraph{Bertotti-Robinson Solutions}\ \\
We now look for AdS$_2 \times S^2$ solutions (\emph{aka} Bertotti-Robinson) which we assume describe the near-horizon geometry of extremal black holes.\footnote{In general, it is not clear whether an  AdS$_2 \times S^2$ is the near-horizon geometry of a full black hole solution. However, this fact is not relevant for our purposes since, if such a extremal black hole solution exists, its horizon will satisfy these constraints.} These solutions are obtained by looking for constant scalar configurations at the horizon denoted by $\bar{\phi}^i$. We write the metric as:
\begin{align}
    \dd s^2 = R_{AdS}^2 \left(\frac{-\dd t^2 + \dd w^2}{w^2} \right) + R^2 \dd \Omega^2\,.
\end{align}
This can be expressed in the language of \eqref{eq:metric_ansatz_flow_eq} through the identification $w = R^2 z$, $e^{2U} = R_{AdS}^2/(R^2 z)^2$, and $e^{2A} = R_{AdS}^2/(R z)^2$. Then, the equations of motion from the effective action \eqref{eq:S_eff_generic_fields} leads to the following equations on the (constant) scalars, $R$ and $R_{\text{AdS}}$: 
\begin{align}
    &\partial_i V  + \frac{1}{R^4} \partial_i V_{BH} = 0\, , \label{eq:scalar_eq_AdS2S2}\\
    &V - \frac{2}{R^2} + \frac{1}{R^4} V_{BH} = 0\, , \label{eq:R_eq_AdS2S2}\\
    &-\frac{2}{R_{AdS}^2} = V-\frac{1}{R^4} V_{BH}\, . \label{eq:RAdS_eq_AdS2S2}
\end{align}

The equations above are algebraic equations that determine $R$, $R_{AdS}$ and $\bar{\phi}^i$. More precisely, the equation for $R_{AdS}$ decouples and it is enough to solve \eqref{eq:scalar_eq_AdS2S2} and \eqref{eq:R_eq_AdS2S2}, from which $R_{AdS}$ follows directly:
\begin{align}
    R_{AdS}^2 = R^2\frac{1}{1-R^2 V(\bar{\phi})}\,.
\end{align}
From these equations, it is clear that when the potential $V(\phi)$ is turned off, the scalar equations of motion of \eqref{eq:S_eff_generic_fields} are extrema of $V_{BH}$ with $R_{AdS}=R$ and $R^2=\frac{1}{2}V_{BH}$. The presence of a cosmological stabilizing potential $V(\phi)$ upsets this picture as there is now a competition between the black hole attractor and the stabilizing potential.

A remark on units, which we use repeatedly below. In \eqref{eq:generic_gravity_action} the scalar kinetic term is $-\frac{1}{2\kappa_4^2}\cdot\frac12(\partial\phi)^2$. The canonically normalised field in reduced Planck units ($\kappa_4=1$) is therefore $\phi/\sqrt{2}$. So for a single scalar with $G=1$, a displacement $\Delta\phi$ corresponds to a geodesic distance $\Delta=|\Delta\phi|/\sqrt2$ in Planck units.

\section{Bottom-up Examples }
\label{sect:bottom_up_EMD_theory}

One's ability to solve the flow equations described in the previous section heavily depends on the details of the scalar potential one chooses. In this section, we solve the equations for toy models with a single scalar field\footnote{While v2 of this paper was in progress, similar ideas were pursued in \cite{Benakli:2026zbm}.} and an exponential coupling to the vector, i.e.\ the following special case of \eqref{eq:generic_gravity_action}:
\begin{equation}
    S=\frac{1}{2\kappa_4^2}\int \dd^4 x \sqrt{-g}\left(R-\tfrac{1}{2} (\partial\phi)^2 -\tfrac{1}{4} e^{a\phi} F^2 - V(\phi)\right) \,. 
\end{equation}
This is a model for the long range forces in our universe if we assume that the fine-structure constant is the vev of a stabilised field.  Such simple models are distinguished by different choices of $V$ and coupling to the vector field. We keep the coupling exponential and discuss a few choices of $V$. Hence for all these models the black hole potential \eqref{eq:BH_potential_generic} is 
\begin{align}\label{eq:BHpotential}
    V_{BH}(\phi)=\frac{1}{2}\left(e^{-a\phi}Q^2 + e^{a\phi}P^2\right)\, .
\end{align}
Expanding around the vacuum, which we for simplicity take at $\phi=0$ (by a simple rescaling of $F$ we can discuss the general situation), we get
\begin{equation}
V(\phi) = \frac{1}{2}m^2\phi^2 +\mathcal{O}(\phi^3)\,,
\end{equation}
where we assumed a Minkowski vacuum.\footnote{In practice, this simply means that we look at black holes whose radii are parametrically smaller than the Hubble scale.}  
Hence, the first model we discuss in most detail has
\begin{equation}
 \text{Model 1:} \quad V(\phi)=\frac{1}{2}m^2\phi^2\,.   
\end{equation}
This has been discussed in earlier literature, in particular in \cite{Horne:1992bi}, but see also  \cite{Gregory:1992kr,Anabalon:2013qua} for related discussions. 

We next consider scalar stabilization in AdS. The negative cosmological potential introduces a qualitative difference from the Minkowski model. Indeed, positivity of $V_{BH}$ implies, through \eqref{eq:R_eq_AdS2S2}, the general horizon constraint
\begin{equation}
    V(\phi_h)<\frac{2}{R^2}\,.
\end{equation}
For the quadratic Minkowski potential, this forces the scalar value at the horizon towards its vacuum value as $R\to\infty$. Therefore, parametrically large BHs can never displace the scalar at the horizon. An AdS minimum, however, has a surrounding region with $V<0$, which this constraint does not exclude even for arbitrarily large horizons. To investigate the resulting scalar displacements, we consider two AdS potentials:
\begin{align}
&     \text{Model 2:} \quad V(\phi)=m^2\left(\cosh(\phi)-2\right)\,.  \,,\\
&     \text{Model 3:} \quad V(\phi)=\frac{2m^2}{a^2}(e^{2a\phi} -e^{a\phi})\,.
\end{align}
Models 2 and 3 have both AdS vacua with AdS length proportional to $m^{-1}$ and mass of the scalar proportional to $m$. This is similar to the known top-down ``non-rigid'' compactifications, in the sense that the mass of the stabilized moduli is always of the same order as the cosmological scale. The difference between Models~2 and~3 lies in their asymptotic behaviour. In Model~2, the potential grows without bound as one moves away from the vacuum. By contrast, Model~3 exhibits asymptotics that are much more in line with string-theoretic expectations: the potential vanishes at ``weak coupling" where $e^{\phi}$ vanishes and blows up in the other limit. This is the generic behaviour of flux potentials in the asymptotic regions of moduli space, where they are sums of exponentials of the canonically normalised saxions \cite{Ooguri:2018wrx,VanRiet:2023pnx}. For this reason \textbf{we regard Model~3 as the toy model representative of string compactifications}, and in particular of the top-down model of section~\ref{sec:topdown}. 

For the three models, we construct the Bertotti-Robinson AdS$_2\times S^2$ solutions, which we think of as the near-horizon geometry of extremal black holes in these theories. Though it is worth understanding better, we do not prove that a given AdS$_2\times S^2$ solution is the endpoint of an attractor flow from the asymptotic vacuum. We postpone such a study to future work and take it as an assumption for now. Note, however, that \emph{if} an extremal black hole with these charges exists, its horizon must satisfy the Bertotti-Robinson equations. We find these solutions perturbatively for the case of ``large'' and ``small'' black holes (compared to the inverse mass of the scalar). Moreover, for model 1 we also solve for the full extremal (large) black hole solution perturbatively, and we comment on its relation with the weak gravity conjecture.
 
\subsection{Model 1: mass term}\label{sec:model1}

For understanding black hole solutions in this setup, two length scales are crucial: the inverse scalar mass $m^{-1}$ and the BH radius $R$. We think of the BH radius as a good proxy for the typical curvature scale at the horizon: the bigger the black hole, the lower the horizon curvature. We now solve the flow equations in two limits
\begin{align}
& \text{Large BHs}:\quad mR \gg 1\,, \nonumber\\
& \text{Small BHs}:\quad mR \ll 1\,.   \nonumber
\end{align}
For large black holes we expect the scalar to be less affected by the black hole and remain close to its cosmological minimum, since the curvature scale at the horizon is very low compared to the scalar mass. In this limit, the black hole is effectively a Reissner-Nordstrom (RN) black hole. In the other limit, the dilaton is effectively massless and we can see the usual attractor mechanism taking place. The black hole solution would be the well-known dilatonic black hole, where the dilaton value at the horizon is set by the black hole charges.

In section \ref{sec:LBHtoymodel} we will solve the equations of motion perturbatively in $\epsilon = (m R)^{-2}$ to see how the RN black hole is modified by the presence of the scalar. Black holes that are small enough should affect the scalars significantly. In Section \ref{sec:SBHtoymodel} we will therefore solve the equations of motion perturbatively in $\delta = (m R)^2$.

\subsubsection{Large Black Holes}\label{sec:LBHtoymodel}

For (extremal) large black holes, we solve the EOMs arising from \eqref{eq:S_eff_generic_fields} perturbatively in $\epsilon=\frac{1}{(mR)^2}$ in appendix \ref{app:BH_model_1}. We summarize here the main intuition and results, and keep only the derivation (general solution with integration constants and how they are fixed) in the appendix.
At 0th order, the dilaton is completely frozen at $\phi=0$ and our theory effectively reduces to pure Einstein-Maxwell theory, where the unperturbed solution is the extremal Reissner-Nordstrom (RN) black hole. Expanding the dilaton $\phi(z)$ and metric functions $U(z)$, $A(z)$ perturbatively in $\epsilon$, one can easily show that the dilaton profile is given by 
\begin{align}
    \phi(z) &= -\frac{a}{2}R_0^2e^{4U_0-4A_0}\left(P^2-Q^2\right)\epsilon\, .
\end{align}
where $U_0$ and $A_0$ are the metric functions for the extremal RN black hole:
\begin{align}
    e^{-U_0(z)} = 1-Mz\, , \quad e^{-2A_0(z)} = z^2
\end{align}
with $M=R_0=\frac12\sqrt{P^2+Q^2}$. This means that, at leading order, $|\phi|$ is bounded from above by its value at the horizon:
\begin{align}
    \phi_h &= -2a \frac{P^2-Q^2}{P^2+Q^2} \epsilon \, .
\end{align}
Regardless of how large we make the charges, $\phi_h$ can never blow up. As expected, a black hole with large radius compared to $\frac{1}{m}$ is not able to displace the dilaton far away from the minimum of its cosmological potential. As we make the black hole smaller, there will be a point where perturbation theory breaks down and it will be roughly at that scale, where $R \sim \frac{1}{m}$, that the competition between the black hole potential and the cosmological potential can appear.\\

We now turn to describing how the geometry is affected by the presence of the scalar. As explained in appendix \ref{app:BH_model_1}, the metric functions at leading order can be written as $U=U_0+\epsilon U_1$, $A=A_0+\epsilon A_1$, where
\begin{align}
    U_1 &= \frac{a^2\left(P^2-Q^2\right)^2}{16R_0^4}\frac{1}{1-R_0z} \left(\frac{59}{30}
    -4 \ln(1-R_0z) + \frac{P_U(1-R_0z)}{(1-R_0z)^3}\right)\, , \label{eq:U1_RN}\\
    A_1 &= \frac{a^2 \left(P^2-Q^2\right)^2}{16R_0^4} \, \frac{1}{R_0 z}\left(- \frac{52}{15}  + 
    4\ln(1-R_0z)
    +\frac{P_A(1-R_0z)}{(1-R_0z)^4} \right)\, , \label{eq:A1_RN}
\end{align}
and where 
\begin{align}
    P_U(x) &= \frac{3}{2}x^4 
    -\frac{9}{2}x^2+\frac{6}{5}x-\frac{1}{6}\, ,\\
    P_A(x) &= -x^5 + \frac{13}{2}x^3 -\frac{27}{10}x^2 +\frac{23}{30}x -\frac{1}{10}\, .
\end{align}
The constants of integrations obtained from solving the EOMs can be fixed by enforcing the Hamiltonian constraint \eqref{eq:Hamiltonian_constraint_generic_fields}, asymptotic flatness and extremality (perturbatively in $\epsilon$). 

The above solution has the following Bertotti-Robinson solution as its near-horizon geometry:
\begin{align}
    R_{AdS}^2 &= R_0^2\left(1+\epsilon a^2\left(\frac{P^2-Q^2}{P^2+Q^2}\right)^2+ \mathcal{O}(\epsilon^2)\right) \, ,\label{eq:AdS2xS2_R_AdS}\\
    R_{S}^2 &= R_0^2\left(1-\epsilon a^2\left(\frac{P^2-Q^2}{P^2+Q^2}\right)^2+ \mathcal{O}(\epsilon^2)\right) \label{eq:AdS2xS2_R_S}\, ,\\
    \phi &= -2a\frac{P^2-Q^2}{P^2+Q^2}\epsilon  + \mathcal{O}(\epsilon^2) \, .\label{eq:AdS2xS2_phi}
\end{align}
As a nice cross-check, one can verify that the same exact profiles can be obtained from solving the near-horizon equations in \eqref{eq:scalar_eq_AdS2S2}-\eqref{eq:RAdS_eq_AdS2S2} perturbatively in $\epsilon$. Using these equations, we can derive the ADM mass $M$ to first order:
\begin{align}
    M = \lim_{z \to 0^-} U'(z) = M_{RN}\left(1 - \frac{1}{m^2M_{RN}^2}\frac{a^2\left(P^2-Q^2\right)^2}{160 R_0^4}\right)\,,
\end{align}
This shows that the mass decreases when ``integrating in" the dilaton, which is consistent with the Weak Gravity Conjecture \cite{Arkani-Hamed:2006emk, Cheung:2018cwt}. 

Alternatively we can write an EFT expansion that introduces the dilaton as higher derivative corrections. In order to find the Wilson coefficients, we consider the equation of motion for $\phi$:

\begin{align}
    \square \phi &= \frac{a}{4}e^{a\phi}F^2 + m^2\phi\, .
\end{align}

At energy scales much smaller than $m$, the above equation can be solved recursively:

\begin{align}
    \phi = -\frac{1}{m^2}\frac{a}{4}F^2 - \frac{1}{m^4}\frac{a}{4}(\square-\frac{a^2}{4}F^2)F^2 + \ldots
\end{align}

Plugging this approximate solution in the action yields

\begin{align}
    \frac{2\kappa_4^2 \mathcal{L}}{\sqrt{-g}} \supset
    \frac{a^2}{32\, m^2}\,(F^2)^2
    + \frac{a^2}{32\, m^4}\,F^2\square F^2
    -\frac{a^4}{128\, m^4}\,(F^2)^3
    + \mathcal{O}(m^{-6})\,.
\end{align}
These terms give meaningful contributions up to the cutoff $\Lambda_{UV}=m$.

To summarize, as expected we find that large black holes have suppressed corrections related to the dynamics of the heavy scalar field. The small corrections are consistent with the WGC and imply small scalar excursions, in line with \cite{Green:2006nv}. 

\subsubsection{Small Black Holes}\label{sec:SBHtoymodel}

We now discuss small black holes in perturbation theory using the small parameter $\delta = (mR)^2$.
At leading order we have a vanishing scalar potential and extremal solutions in this limit exhibit the attractor mechanism. Using equation \eqref{eq:scalar_eq_AdS2S2} with $V=0$, the value of the scalar at the horizon, $\phi_h$ is fixed in terms of the black hole charges:
\begin{align}\label{eq:dilatonicbh}
    \partial_i V_{BH} &= 0 \implies e^{2a\phi_h}=\frac{Q^2}{P^2}\, ,
\end{align}
independently of its asymptotic value. Moreover, the radius of the black hole is given by
\begin{align}\label{eq:dilb}
    R_0^2 &= \frac{1}{2}V_{BH}(\phi_h) = \frac{1}{2}|QP|\, .
\end{align}
By choosing a large electric charge $Q$ for the black hole, the scalar is dragged by $|a\phi_h|=\ln|Q/P|$, i.e.\ a geodesic distance $\Delta=\ln|Q/P|/(\sqrt2|a|)$ ($P$ cannot be sent to zero whilst preserving charge quantization), which can be made arbitrarily large. This is the massless behaviour recalled in section~\ref{sec:flows}.

In order to find the corrections due to the scalar potential, it is enough to solve for the AdS$_2 \times S^2$ horizon perturbatively in $\delta = (mR_0)^2$. We find
\begin{align}
    \phi &= D_0 \left[1-\delta \frac{1}{2a^2} + \mathcal{O}(\delta^2)\right]\, ,\\
    R^2 &= R_0^2 \left(1+\frac{1}{4}\delta D_0^2 + \mathcal{O}(\delta^2)\right)\, ,\\
    R_{AdS}^2 &= R_0^2 \left(1+\frac{3}{4}\delta D_0^2 + \mathcal{O}(\delta^2)\right)\, ,
\end{align}
where $D_0 \equiv \phi_h$ is the attractor value \eqref{eq:dilatonicbh}. It is therefore clear that for any finite value of $R_0$ (and for small $m$), the presence of the mass term only modifies the distance traversed in field space by a small quantity. Another way to see this is as follows. One of the features of the attractor mechanism in eqs. \eqref{eq:dilatonicbh} and \eqref{eq:dilb} (and of attractor mechanisms more generally) is that one can always rescale the charges by a constant (whilst being careful to preserve charge quantization) and have the same attractor point: this means that for a fixed value of $m$, we can make $R_0$ small, leaving $D_0$ invariant. In this limit it is also clear that the shift of the horizon value $\delta D_0$ is also small:

\begin{align}
    \delta D_0 = \underbrace{\frac{m^2}{2a^2} D_0}_{fixed}\; R_0^2 \, ,
\end{align}
so even though $D_0$ is very large, the shift is small and leading order perturbation theory can be trusted. Physically, our intuition is realized: the scalar potential stabilizes $\phi$ to 0 at infinity, but the small black hole is powerful enough to drag the scalar to the would-be attractor value, with a small $\delta = m^2 R_0^2$ correction. 

\subsection{Model 2: cosh potential}
We now consider the potential:
\begin{equation}\label{eq:coshpot}
V(\phi)=m^2\left(\cosh(\phi)-2\right)\,.    
\end{equation}
This potential has an AdS vacuum at $\phi =0$ with $V(0)=-m^2$, i.e.\ $L_{AdS_4}^2=6/m^2$ in the normalisation of \eqref{eq:generic_gravity_action}, and scalar mass $m$ in the vacuum. Let us further fix the exponential coupling to the Maxwell vector as $a=1$ for simplicity. The Bertotti-Robinson \eqref{eq:scalar_eq_AdS2S2}, \eqref{eq:R_eq_AdS2S2}equations can be combined to get
\begin{align}
    \frac{Q^2}{R^2} &= 2(1+m^2R^2)e^{\phi} - m^2R^2 \label{eq:Model2_Q_eq}\\
    \frac{P^2}{R^2} &= 2(1+m^2R^2)e^{-\phi} - m^2R^2 \label{eq:Model2_P_eq}
\end{align}
where $\phi$ is the horizon value fo the scalar. Using positivity of $Q^2$ and $P^2$, this leads to the following bound on the scalar value at the horizon:
\begin{align}
    |\phi| \leq \ln\left(2\left(1+\frac{1}{m^2R^2}\right)\right)\, ,
\end{align}
where the bound is saturated for purely electric or purely magnetic solutions. For large BHs ($mR\gg 1$), the scalar value is bounded above by $\ln(2)$: large BHs cannot displace the scalar arbitrarily far from its minimum, in line with intuition. For small BHs ($mR\ll 1$), the bound grows logarithmically, so the scalar field can potentially be displaced significantly (i.e. travel a large distance from its minimum in planck units). While this confirms our intuition, it is worth showing a small dyonic example for illustration: consider fixed $P\neq 0$, and vary $Q$ to get small or large BHs. Equation \eqref{eq:Model2_P_eq} leads to
\begin{align}
    e^{\phi} &= \frac{2R^2(1+m^2R^2)}{P^2+m^2R^4}
\end{align}
which we can then plug into \eqref{eq:Model2_Q_eq} to get 
\begin{align}
    \frac{Q^2}{R^2} &= \frac{4R^2(1+m^2R^2)^2}{P^2+m^2R^4} - m^2R^2
\end{align}
Then,
\begin{itemize}
    \item For large BHs we scale $m|Q| \to \infty$ and obtain
    \begin{align}
        m^2R^2 &= \frac{m|Q|}{\sqrt{3}} + \mathcal{O}(1)\, , \quad e^{\phi} = 2\left(1+\frac{\sqrt{3}}{m|Q|}+\mathcal{O}((m|Q|)^{-2})\right)
    \end{align}
    as well as an AdS$_2$ radius given by
    \begin{align}
        m^2 R_{AdS}^2 &= \frac{4}{3} + \mathcal{O}((m|Q|)^{-1})
    \end{align}
    so even though the $S^2$ radius grows with $m|Q|$, the AdS$_2$ radius stays fixed: this is an example of inverse 2d scale separation. Note also here that for arbitrarily large BHs, the scalar is displaced a distance $\Delta=\frac{\ln(2)}{\sqrt{2}}$ in Planck units. The reason why the displacement does not vanish as $R\to\infty$, contrary to the Minkowski case, is that in AdS the electromagnetic energy density at the horizon of an extremal black hole does not go to zero for large horizons. Indeed \eqref{eq:R_eq_AdS2S2} gives $V_{BH}/R^4=2/R^2-V(\phi_h)\to-V(\phi_h)=\frac34m^2$, the analogue of $Q\sim R^2/L_{AdS}$ for extremal RN-AdS black holes. The horizon thus always carries an energy density of the order of the vacuum energy, which is enough to shift the scalar by an $\mathcal{O}(1)$ amount, and the AdS$_2$ throat has a fixed size $R_{AdS}\sim 1/m$. The shift is finite and independent of the size of the black hole.

    \item For small BHs, we need instead to take $m|Q|\to 0$. These solutions only exist when $m^2P^2 < 4$, and we get
    \begin{align}
         m^2R^2 &= \frac{m^2|QP|}{\sqrt{4-m^2P^2}} + \mathcal{O}((m|Q|)^2)\, ,\\
         e^{\phi} &= \frac{2|Q|}{|P|}\frac{1}{\sqrt{4-m^2P^2}}+\mathcal{O}((m|Q|)^2)\, .
    \end{align} 
    There, we have a small BH with $\phi$ travelling a parametrically large distance ($\phi \to -\infty$). If we additionally take the limit of small magnetic charge ($m|P|\ll 1$), we recover the massless scalar case, see \eqref{eq:dilatonicbh}-\eqref{eq:dilb}. The AdS$_2$ length scale can be found in this limit as well:
    \begin{align}
        m^2R_{AdS}^2 &= m^2 R^2\frac{4}{4-m^2P^2} +\mathcal{O}((m|Q|)^2)\, ,
    \end{align}
    so $R_{AdS}$ is of the same order as the $S^2$ radius $R$.
\end{itemize}
To summarize, only small black holes ($mR\ll 1$) can produce parametrically large scalar displacements in this model, in line with expectations. Note also that for the equal charge case ($Q=P$), the scalar remains at its cosmological minimum regardless of BH size: the displacement of the scalar that we are interested in is not a generic feature for all BH charges. 

Now let us turn to a more phenomenologically oriented question: what would happen in our universe if there were scalars with mass given by the hubble scale, as in the example above? If we consider the ratio of the diameter of Sagittarius A* with the Hubble scale we get an $m^2R^2$ of the order of $10^{-32}$.\footnote{Using a value of $\sim 10^{10}$m for the diameter of the black hole, and $\sim 10^{26}$ m for the Hubble scale.} This would mean a possible field displacement $\Delta\phi=\ln\big(2(1+10^{32})\big)\approx 74$, i.e.\ a geodesic distance $\Delta\approx 52$ in Planck units, for what is considered to be a huge black hole. The point is that, in units of the inverse scalar mass (here the Hubble length), Sagittarius A$^*$ is a \emph{small} black hole.

\subsection{Model 3: Double exponential}

This last toy model is much more in line with expectations from string theory:
\begin{equation}
V(\phi)=\frac{2m^2}{a^2}(e^{2a\phi} -e^{a\phi})\,, 
\end{equation}
where $a$ is the same constant as in the black hole potential. The asymptotics of the potential is that it vanishes at weak coupling and blows up at strong coupling (if we think of $e^{a\phi}$ as a coupling). The cosmological minimum is at
\begin{equation}
e^{a\phi}=\frac{1}{2}\qquad \text{where}\qquad V=-\frac{m^2}{2a^2}\, 
\end{equation}
and the mass of the scalar is equal to $m$. Therefore, both the mass and the cosmological constant are given by the scale $m$. As in model 2, one can combine the first two BR equations to get expressions for each charge:
\begin{equation}\label{eq:model3_charges}
P^2=2R^2e^{-a\phi}\left(1-\frac{m^2R^2}{a^2}e^{a\phi}(3e^{a\phi}-2)\right)\,,\qquad Q^2=2R^2e^{a\phi}\left(1+\frac{m^2R^2}{a^2}e^{2a\phi}\right)\,,
\end{equation}
so a Bertotti-Robinson solution exists whenever the right-hand side of the first equation is non-negative. This in turn places a bound on $e^{a\phi}$:
\begin{align}
    0 < e^{a\phi} \leq \frac{1 + \sqrt{1+\frac{3a^2}{m^2R^2}}}{3}\, ,
\end{align}
where the upper edge of this window corresponds to a purely electric solution\footnote{Interestingly, small BHs in the purely electric solution actually drag the scalar towards positive values where the potential is blowing up ($e^{a\phi} = \frac{\sqrt{3}a}{3mR}+\mathcal{O}(1)$), even though the potential is asymptotically flat in the other direction.}. Note the asymmetry of the range, where $\phi \to -\infty$ is (naively) allowed regardless of the size of the BH, but the other direction $\phi\to\infty$ is not allowed unless the BH is small. This is the main qualitative difference of model 3 compared to model 2. \\

For this toy model, we illustrate two charge regimes with different physics below: first, fixing $P$ and varying $Q$, and then doing the opposite.

\paragraph{Fixed $P$, varying $Q$}\ \\

We solve the BR equations numerically for $\phi$ and $R$ at fixed $P$, varying $Q$ (see figure \ref{fig:fixedP} below) to get an idea of the space of solutions.
\begin{figure}[H]
    \centering
    \includegraphics[width=0.7\linewidth]{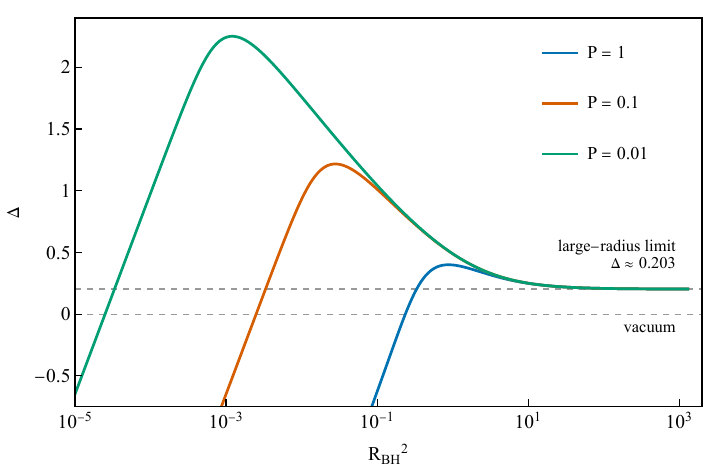}
    \caption{Behavior of the (signed) scalar distance $\Delta=\frac{\phi_h-\phi_{\rm vac}}{\sqrt{2}}$ as the black hole size is increased (by varying $Q$ between $10^{-3}$ and $10^{3}$) for different values of $P$, with $m=a=1$.}
    \label{fig:fixedP}
\end{figure}
As previously, we see that in the large BH limit the scalar reaches a constant value\footnote{$e^{a\phi}=\frac{2}{3}$, since at very large $Q$ one can essentially neglect $P$ and we are again in an approximately electric solution} that is an $\mathcal{O}(1)$ distance from its cosmological minimum in planck units. On the other end of the figure, we see that $\phi \to -\infty$, indicating a large distance travelled for small BHs. There, we recover again the ``massless scalar'' physics, where $\Delta=\frac{1}{\sqrt{2}a}\ln(\tfrac{4R^2}{P^2})$\footnote{This scaling of the scalar distance with the log of the BH size is why figure \ref{fig:fixedP} exhibits linear growth in the small BH region.}. The zero-size endpoint $Q=0$ itself is singular and is not part of our analysis. Moreover, the bump in figure \ref{fig:fixedP} shows the transition from approximately massless attractor behaviour (where $\phi$ increases with the BH size) to the regime where the cosmological potential can no longer be neglected.

\paragraph{Fixed $Q$, varying $P$}\ \\
Here we can see the effect of the flat direction of the potential very clearly: indeed, one can find large BHs (mostly made up of magnetic charge since we keep $Q$ order 1) which displace $\phi \to -\infty$. Indeed, the BR equations at large $P$ can be solved to get 
\begin{align}
    e^{a\phi} &= \frac{|Q|}{|P|}\sqrt{1+\frac{m^2Q^2}{a^2}} + \mathcal{O}(P^{-2})\\
    R^2 &= \frac{|PQ|}{2\sqrt{1+\frac{m^2Q^2}{a^2}}} + \mathcal{O}(1)
\end{align}
We therefore see how an arbitrarily large BH can still drag the scalar far from its minimum, which makes sense given that it is dragging the scalar towards the flat direction of the potential. Moreover, the AdS$_2$ scale is proportional to the $S^2$ scale as
\begin{align}
    R_{AdS_2}^2 = \frac{R^2}{1+\frac{m^2Q^2}{a^2}} + \mathcal{O}(P^{-1})\, . 
\end{align}  
Therefore both $S^2$ and $AdS_2$ radii squared grow with $|P|$. Combining the two expressions above, the horizon value at leading order is $e^{a\phi_h}=Q^2/(2R^2)$, independently of $m$. So the scalar distance
\begin{equation}
\Delta=\frac{|\phi_h-\phi_{\rm vac}|}{\sqrt2}=\frac{\sqrt2}{|a|}\ln\frac{R}{|Q|}\,,
\end{equation}
grows logarithmically with the size of the black hole, exactly as for a massless scalar. It is unbounded only as $R\to\infty$ and finite for any finite black hole. The reason is that the potential and its gradient vanish exponentially along the flat direction. In other words, measured with respect to the mass of the modulus \emph{at its horizon value}, these black holes do not become larger as $P$ grows, even though $mR\gg1$ in terms of the vacuum mass. Hence the vacuum mass alone does not decide whether a black hole is ``large''. Note that this requires a specific charge regime (large $P$ at fixed $Q$): at fixed $P$ and large $Q$ the displacement stays $\mathcal{O}(1)$, as shown above.

\subsection{Lessons learned from bottom-up models}
The near-horizon solutions illustrate how scalar displacements depend on both the vacuum mass and the shape of the potential. For the quadratic Minkowski potential, large black holes ($mR\gg1$) keep the scalar close to its minimum. Model 2 shows that an AdS minimum instead permits $\mathcal{O}(1)$ displacements in Planck units at large radius. In Model 3, the potential has a flat direction, allowing suitable charge scalings to produce arbitrarily large horizons with unbounded scalar excursions, growing logarithmically with $R$ as in the massless case. Small black holes ($mR\ll1$) can also produce large excursions; for scalars with masses of order the Hubble scale, this regime includes astrophysical black holes. We will now establish such solutions in top down models which can be described by EFTs in four-dimensions. A necessary condition for that description is that the 4d AdS vacuum is scale separated.

\section{Top-down Examples}\label{sec:topdown}

We now turn to explicit string compactifications to investigate whether the behaviour identified in the bottom-up models can also occur in a top-down setting. Of particular interest is the mechanism
illustrated by Model~3: even black holes with $mR\gg1$ can produce unbounded scalar excursions along an asymptotic direction where the potential approaches zero from below. Therefore, the vacuum mass alone is insufficient to determine the possible displacement: the interplay between the black hole charges and the full moduli potential needs to be analyzed in a concrete compactification.

Since moduli stabilisation is best understood in either IIA or IIB string theory we aim to construct charged black hole solutions in such theories. Concerning the situation in IIB flux compactifications, the first step in this direction was made in \cite{Danielsson:2006jg}, where they recovered similar physics to that of section \ref{sect:bottom_up_EMD_theory} but many subtleties appear due to the orientifolding which was not done carefully. In particular, in order to realize such black holes as D3 branes wrapped on 3-cycles in the Calabi-Yau, one must be careful about the presence of fluxes and the possibility of Freed-Witten effects. We review the construction of \cite{Danielsson:2006jg}  and discuss how one can improve the situation towards a full-fledged embedding in string theory in Appendix \ref{App:IIB}. Beyond the technical difficulties there are in constructing this embedding, one is still ignoring all of the K\"ahler sector of moduli space. This is a big approximation, since the no-scale Minkowski backgrounds that we consider will certainly be altered after K\"ahler moduli stabilisation is taken into account. Whether or not the black hole solutions will be strongly altered by this depends on whether the K\"ahler moduli couple to the vectors after full moduli stabilisation. Famous suggestions on how to achieve full moduli stabilisation in this context are based on the so-called KKLT scenario \cite{Kachru:2003aw} (or its modern follow-ups \cite{Demirtas:2021nlu}) or the Large Volume Scenario \cite{Balasubramanian:2005zx}. Unfortunately these scenarios lack a full top-down understanding and as a consequence their consistency is still being debated, see for instance \cite{Lust:2022lfc, Bobev:2023dwx}.

If we instead turn to IIA flux vacua we can be much more concrete. The IIA vacuum will have a negative cosmological constant and the mass of the scalars is order one in AdS units. This example features black holes horizons whose existence depends entirely on a combination of the cosmological and black hole potential. 
In particular, we consider the scale separated  flux vacua in massive IIA string theory obtained from CY-O6 reductions with Romans mass $F_0$, $H_3$ flux and unbounded $F_4$ fluxes \cite{Derendinger:2004jn, DeWolfe:2005uu, Camara:2005dc}, based on the formalism of \cite{Grimm:2004ua}. 

In this context, vectors are only generated in the presence of even two-cycles from reducing $C_3$ over these cycles.  Before we continue with the details we mention that these vacua are still being contested as fully consistent mainly because they have only been derived in the limit in which the O-planes are smeared \cite{Lust:2004ig,Acharya:2006ne} as this limit coincides with the $N=1$ EFT in 4d \cite{Grana:2006kf}. Establishing the consistency of the vacua beyond this is currently being debated \cite{Junghans:2020acz,Marchesano:2020qvg, Baines:2020dmu, Cribiori:2021djm,VanRiet:2023pnx, Emelin:2024vug, Montero:2024qtz}.  

\subsection{The $\mathbb{T}^6/\mathbb{Z}_4$ Orbifold}

We follow references \cite{Ihl:2006pp, Ihl:2007ah} for a concrete example based on the orbifold $\mathbb{T}^6/\mathbb{Z}_4$. This orbifold has an even two-cycle and hence a single vector: we can contemplate simple charged black holes. Since the charges come from $C_3$, the electric and magnetic charges correspond to D2 branes wrapping the even 2-cycle and D4 branes wrapping the dual 4-cycles. 

In appendix \ref{app:T6_Z4_orbifold_details}, following the conventions of \cite{Ihl:2007ah}, we review the 4d $\mathcal{N}=1$ EFT that arises from compactifying type IIA supergravity on the $\mathbb{T}^6/\mathbb{Z}_4$ orbifold. Instead, here we just directly present the effective field theory. The bosonic part of the $\mathcal{N}=1$  supergravity action is\footnote{Note the different normalisation of the Planck mass with respect to the previous section. }
\begin{equation}
    S= \int \sqrt{-g}\left(\frac12 R   - K_{ I \bar J } \partial \phi ^I \partial\bar{\phi} ^{\bar J} - V - \frac14 (\text{Re} f_{\alpha \beta}) F^\alpha_{\mu\nu}  F^{\beta\mu\nu} - \frac18 ( \text{Im} f_{\alpha \beta}) \epsilon^{\mu\nu\rho\sigma}F^{\alpha}_{\mu\nu} F^{\beta}_{\rho\sigma} \right)\,, \label{eq:4d_SUGRA_action}
\end{equation}
with the scalar potential 
\begin{equation}\label{eq:scalpot}
    V= e^K \left(K^{I\bar J} D_ I W D_{\bar J } \bar{W} - 3 |W|^2\right) + \frac12 (\text{Re} f) ^{- 1 \alpha \beta}D_ \alpha D_ \beta\,.
\end{equation}
Here, $K_ { I \bar J } = \partial _ I \partial _ { \bar J } K$, and $D_ I W = \partial _ I W + ( \partial_I K) W$ is the K\"ahler covariant derivative. For the case at hand the D-terms in the potential $V$ vanish. One can demonstrate the absence of $D$-terms as long as the internal geometry is CY \cite{Ihl:2006pp,Ihl:2007ah}. 

Our orbifold model, in the untwisted sector\footnote{We commit a common sin and ignore the twisted sector.}, is described by 6 complex scalar fields:
\begin{equation}
\text{complex scalars:}\quad \{N_1, N_2, t_a\,,\quad a=1\ldots 4\}\,.    
\end{equation}
The $t_a$ correspond to complexified K\"ahler deformations, one of the $N$'s is a complexified complex structure deformation and the other $N$ corresponds to the complexified four-dimensional dilaton field. In terms of the 12 real fields we write
\begin{equation}
t_a = u_a + i\nu_a\qquad N_{1,2} = \frac{\xi_{1,2}}{2} +ie^{-D}Z_{1,2} \,,    
\end{equation}
with the $Z$'s written in terms of a single real function $U$ through:
\begin{equation}
    Z_1 = \frac{1}{2 \sqrt{U}}\left(\frac12 + U\right)\,,\qquad Z_ 2  = \frac{1}{2 \sqrt{U}}\left(\frac12 - U\right)\,. 
\end{equation}
So the 12 real fields are
\begin{equation}
\text{real scalars:}\quad \{u_a, \nu_a, D, U, \xi_1, \xi_2\,,\quad a=1\ldots 4\}\,.    
\end{equation}
The details of the scalar kinetic term can be found in appendix \ref{app:T6_Z4_orbifold_details}. Out of the 12 scalars, 11 can be stabilized by turning on fluxes as explained in appendix \ref{app:flux_contribution}. Importantly, the addition of these fluxes has no effect on the kinetic terms of the four-dimensional fields nor does it change the K{\"a}hler potential. What matters here is the value of the scalar potential at the SUSY minimum:
\begin{equation}
    V_{min}= -\frac{243}{200}\left(p_1^2-p_2^2\right)^2\sqrt{\frac{3m_0^{5}}{5\hat{e}_3^3 \left(\hat{e}_1\hat{e}_2-2 \hat{e}_4^2\right)^3}}  = -\frac{3}{L_{AdS_4}^2}\,,
\end{equation}
in 4d Planck units ($p_1,p_2,m_0$ and $\hat{e}_a$ are flux numbers, defined in appendix \ref{app:flux_contribution}). We have verified this value, the location of the SUSY vacuum (appendix~\ref{app:flux_contribution}) and the spectrum quoted below, by constructing the full potential \eqref{eq:scalpot} from the K\"ahler potential and superpotential of appendix~\ref{app:T6_Z4_orbifold_details} independently of \cite{Ihl:2007ah}. The 4d supergravity approximation can be trusted when $L_{AdS_4}$ is large. Moreover, as was shown in \cite{Apers:2022vfp}, the theory can be expanded to quadratic order around this AdS$_4$ vacuum to find the 12 scalar masses. These are of order 1 in AdS$_4$ units, leading to integer\footnote{Precisely, the 12 conformal dimensions are: 2, 3, 5, 5, 5, 5, 6, 6, 6, 6, 10, 11.} conformal dimensions.\\

It is also important to note that the scalar fields $\nu_a$ parametrize the geometry of the orbifold covering space: the metric on $\mathbb{T}^6$ is \cite{Ihl:2007ah}
\begin{align}
    \dd s^2= & \nu_1\left(\dd x_1^2+\dd y_1^2\right)+\nu_2\left(\dd x_2^2+\dd y_2^2\right) +\frac{\nu_3}{U}\left(\dd x_3^2+\dd x_3 \dd y_3+\left(\frac{1}{4}+U^2\right) \dd y_3^2\right) \nonumber \\
    & -2 \nu_4\left(\dd x_1 \dd x_2+\dd x_1 \dd y_2-\dd y_1 \dd x_2+\dd y_1 \dd y_2\right)\,.
\end{align}

Finally, in order to compute the black hole effective potential we need the expressions for the gauge kinetic term. In our specific model there is only one vector and it couples only to the complex scalar $t_3$ via
\begin{equation}
f=-it_3=\nu_3 -iu_3 \,.
\end{equation}
This leads (using \eqref{eq:BH_potential_generic} adapted to the normalisation of \eqref{eq:4d_SUGRA_action}, i.e.\ $V_{BH}=\frac12 P^2\,\text{Re}f+\frac{1}{2\,\text{Re}f}\left(Q-P\,\text{Im}f\right)^2$, see appendix~\ref{app:Stability_BR_solutions}) to the black hole potential
\begin{align}
    V_{BH}(t_3) &= 
    \frac{1}{2\nu_3}\left(Q + P u_3\right)^2 + \frac{1}{2}P^2\nu_3\, . \label{eq:BH_potential_t3}
\end{align}
This simple structure was our main motivation for the choice of orbifold.

\subsection{Bertotti-Robinson Solutions}
When we remove the cosmological moduli stabilisation such that we only have the black hole effective potential, our model has the peculiar property of not having extremal black holes with finite horizon size. Yet, as we now show, thanks to the flux-induced scalar potential extremal black holes with finite horizon sizes do exist\footnote{We are somewhat sloppy in our language since we construct AdS$_2\times S^2$ vacua instead of full black hole flow.}. 

For generic charges, the 12 Bertotti-Robinson equations are difficult to solve. However, we note a surprising simplification in the purely electric case with specific flux choices, which is detailed below. There, we find two distinct branches of near-horizon solutions at the same charge, above a minimum charge. Both branches admit arbitrarily large horizons as the charge is increased, but their scalar behaviour differs: the displacement
remains bounded on one branch, while it grows without bound on the other, along a direction where the scalar potential vanishes from below (as in toy model 3).\\

In order to find AdS$_2\times S^2$ vacua, we need to solve the following algebraic equations for all 11 scalars and $R$:
\begin{align}
    &\partial_{\Phi_A} V   = 0\, , \label{eq:NoBH_eq_AdS2S2_top_down}\\
    &\partial_{t_3} V  + \frac{1}{R^4} \partial_{t_3} V_{BH} = 0\, , \label{eq:t3_eq_AdS2S2_top_down}\\
    &V - \frac{1}{R^2} + \frac{1}{R^4} V_{BH} = 0\, .\label{eq:R_eq_AdS2S2_top_down}
\end{align}
where $\Phi_A=\{u_1,u_2,u_4,\nu_1,\nu_2,\nu_4,D,U,\phi_1\}$, i.e. the 9 scalars which do not enter in the black hole potential (here $\phi_1=p_1\xi_1+p_2\xi_2$ is the combination of the axions $\xi_K$ that appears in the potential, see appendix~\ref{app:flux_contribution}; the orthogonal combination $\phi_2$ is a flat direction and drops out). Note the $\frac{1}{R^2}$ in \eqref{eq:R_eq_AdS2S2_top_down}, as opposed to the $\frac{2}{R^2}$ in \eqref{eq:R_eq_AdS2S2}. This is due to the supergravity normalisation $\frac12\mathcal{R}$ of the Einstein-Hilbert term in \eqref{eq:4d_SUGRA_action}, see appendix~\ref{app:Stability_BR_solutions}.

The AdS$_2$ radius is then given by
\begin{align}
   R_{AdS_2}^2 = R^2\frac{1}{1-2 R^2 V}\, .\label{eq:RAdS2_eq_AdS2S2_top_down}
\end{align}
These equations are difficult to solve in general, so we focus on a specific branch which can almost fully be analyzed analytically, namely the electric black hole case ($P=0$).

When setting the magnetic charge to 0, the BH potential only depends on $\nu_3$ as 
\begin{align}
    V_{BH}(t_3) &= \frac{Q^2}{2\nu_3}\, . \label{eq:electric_BH_potential_v3}
\end{align}
As a (surprising) consequence, it is possible to ``truncate'' the problem of solving the Bertotti-Robinson equation to a slightly easier one. Namely, one can choose to set the scalars $u_a$, $U$ and $\phi_1$ to their cosmological minimum while keeping the $\nu_a$ and $D$ free. This choice solves equation \eqref{eq:NoBH_eq_AdS2S2_top_down} for $\Phi_A\in\{u_a,U,\phi_1\}$, and simplifies the other directions. The problem can then be simplified further by taking some symmetric choice of fluxes:
\begin{align}
    m_0 &= 1\, , \quad m_a=0 \, ,\\
    \hat{e}_1 &= \hat{e}_2 = \alpha N, \quad \hat{e}_3=\hat{e}_4 = N \, ,\\
    p_1 &= \frac{-1-2U_0}{2\sqrt{2U_0}}, \, p_2 = \frac{1-2U_0}{2\sqrt{2U_0}}
\end{align}
where $\alpha > \sqrt{2}$ (so that $\hat e_1\hat e_2>2\hat e_4^2$) and $N>0$ (so that $\hat e_{1,2,3}$ have the same sign as $m_0$). This choice of $p_{1,2}$ satisfies $p_1m_0<0$ and $|p_1|>|p_2|$, gives $U^{vac}=\frac12\frac{p_1+p_2}{p_1-p_2}=U_0$, and normalises $p_1^2-p_2^2=1$, which simplifies all formulas below. (Flux quantization and the tadpole condition of appendix~\ref{app:flux_contribution} restrict the allowed values of $U_0$; we will not need to specify them.). The vacuum values of the scalars with this choice of fluxes are
\begin{align}
    u_a^{vac} &= 0\, ,\quad a=1,2,3,4\,,\\
    U^{vac} &= U_0\,, \\
    \nu_{1}^{vac}=\nu_2^{vac} &= \sqrt{\frac{5}{3}} \frac{\alpha}{\sqrt{\alpha^2-2}} N^{1/2}\,,\\
    \nu_4^{vac}&=\sqrt{\frac{5}{3}} \frac{1}{\sqrt{\alpha^2-2}} N^{1/2}\,,\\
    \nu_3^{vac} &= \sqrt{\frac{5}{3}} \sqrt{\alpha^2-2}\, N^{1/2}\,,\\
    e^{2D^{vac}} &= \frac{27}{10(\alpha^2-2)} N^{-3}\, ,
\end{align}
and the 4d EFT can be trusted in the large $N$ regime (as usual for DGKT setups). Note that the AdS$_4$ length is given in terms of $N$ by 
\begin{align}
    L_{AdS_4}^2 = \frac{200}{81} \sqrt{\frac{5}{3}} \left(\alpha^2-2\right)^{3/2} N^{9/2}
\end{align}
Moreover, with this choice of fluxes\footnote{In fact, one can make a more general choice of fluxes that still allows for this structure - but we present here the simplest case for convenience.} one can algebraically solve all but one Bertotti-Robinson equations for all the scalars that do not stay at the cosmological minimum (the $\nu_a$ and $D$). In practice, we ansatz the form of the $\nu_a$, $D$, and $R$ as functions of a variable $x$ (detailed below), and the last BR equation \eqref{eq:R_eq_AdS2S2_top_down} becomes a particularly simple equation fixing this variable $x$. We ansatz
\begin{align}
    \nu_a &= \left(\frac{3\sqrt{4+x^2}}{5x}\right)^{1/2}  \nu_a^{vac}\, , \quad a=1,2,4\, ,\\
    \nu_3 &= \left(\frac{3x\sqrt{4+x^2}}{5(3\sqrt{4+x^2}-3-x^2)}\right)^{1/2}  \nu_3^{vac}\,, \label{eq:topdown_ansatz_nu3}\\ 
    e^{2D} &= \frac{20x(3\sqrt{4+x^2}-3-x^2)}{27\sqrt{4+x^2}} e^{2D^{vac}}\, , \label{eq:topdown_ansatz_D}
\end{align}
as well as the $S^2$ radius $R$
\begin{align}
    R^2 &= L_{AdS_4}^2 |\tilde{Q}| \frac{81}{200 \left(\alpha^2-2\right) } \sqrt{\frac{3}{10}}  \left(\frac{x^2+4}{x^2\left(9 (x^2 + 2) \sqrt{x^2+4} - 2 x^4 - 18 x^2 - 45\right)}\right)^{1/2} \, ,\label{eq:R2_topdown}
\end{align}
where $\tfrac{3}{2}<x<\sqrt{\tfrac{3}{2}(1+\sqrt{13})}$ (this range is precisely the one that makes $D$ and $R$ real, and the left edge of the range $x=\tfrac{3}{2}$ gives the SUSY vacuum). Note that we rescaled the electric charge $\tilde{Q}=Q N^{-5/2}$ for later convenience. 

The remaining BR equation is \eqref{eq:R_eq_AdS2S2_top_down} and now only depends on fluxes and $x$. It can be succintly written as
\begin{align}
    F(x) = \frac{\alpha^2-2}{|\tilde{Q}|}\, , \label{eq:Fx_eq}
\end{align}
where $F(x)$ is given by
\begin{equation}
F(x)=\frac{\sqrt{2x}\,(1+3y-y^2)(y^2-2y-2)}{y^{3/2}\sqrt{(y+1)(2y-5)}}\,, \quad (y=\sqrt{x^2+4})\, .
\end{equation}
Since the right-hand side of \eqref{eq:Fx_eq} is positive, solutions are restricted to the interval
\begin{equation}
    x_c < x < x_R\,,
    \qquad
    x_c = 12^{1/4}\,,
    \qquad
    x_R = \sqrt{\tfrac32(1+\sqrt{13})}\,.
\end{equation}
Here $x_c$ is the zero of $F(x)$ inside the reality range, while $x_R$ is its upper endpoint, where $F(x)$ also vanishes, see figure \ref{fig:Fx} below.
\begin{figure}[H]
    \centering
    \includegraphics[width=0.7\linewidth]{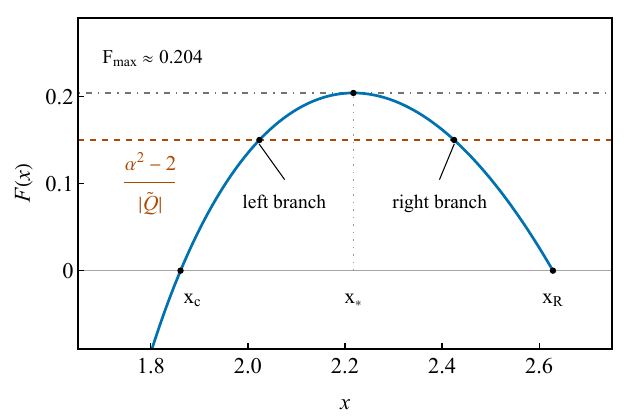}
    \caption{The function $F(x)$ of \eqref{eq:Fx_eq}. It is positive only for $x_c<x<x_R$, where it has a single maximum $F_{\max}\simeq0.204$ at $x_*\simeq2.217$. For $|\tilde Q|>\tilde Q_0=(\alpha^2-2)/F_{\max}$ a horizontal line $(\alpha^2-2)/|\tilde Q|$ intersects it twice, giving the left and right branches of solutions. For $\tfrac32<x<x_c$ one has $F<0$ and there are no solutions.}
    \label{fig:Fx}
\end{figure}

Since $F$ has a maximum, there is a minimal value of the charge such that an AdS$_2\times S^2$ solution exists (within the truncation described above). At that ``minimal BH point'', $x=x_*\simeq2.217$ and all fields are an $O(1)$ number\footnote{it is not possible to have closed form expressions, only numerical ones since one has to find roots of high degree polynomials.} away from the vacuum values. The size of the $S^2$ at this point is 
\begin{align}
    R^2 &\approx 0.729 L_{AdS_4}^2
\end{align}
Moreover, the corresponding minimal charge is $\tilde{Q}_0\approx 4.9(\alpha^2-2)$, i.e. an order 1 number\footnote{This is the reason why $\tilde{Q}=QN^{-5/2}$ was defined earlier - in the $N\to\infty$ limit, the minimal charge $Q$ goes like $Q_0=N^{5/2}\tilde{Q}_0$ so one necessarily has to have a very large $Q$-charge to have solutions, whereas if one uses $\tilde{Q}$ the minimal charge is $N$-independent.}. Within this truncation, we therefore cannot find AdS$_2 \times S^2$ vacua with arbitrarily small $S^2$.
    
As one increases the charge above this critical value, there exists two branches of solutions: one where $x$ is decreasing as $Q$ increases, and another where the opposite happens. We call these two branches the ``left'' branch and the ``right'' branch for obvious reasons. We now discuss the two branches separately. A full black hole flow from the vacuum to the AdS$_2\times S^2$ would presumably pick out one of the two branches as near horizon solutions,

\paragraph{The left branch:}\ \\
On the left branch, as $\tilde{Q}$ is increased one converges to the value of $x$ where $F$ has a zero: $x_c=12^{1/4}\approx 1.86$. At this point, all fields are simply an $\mathcal{O}(1)$ distance away from their cosmological minimum\footnote{For instance $\nu_{1,2,4}/\nu_{1,2,4}^{vac}\simeq0.94$, $\nu_3/\nu_3^{vac}\simeq1.33$ and $e^{2D}/e^{2D^{vac}}\simeq0.87$ at $x_c$.}, and the BH is arbitrarily large: $R^2 \propto N^2 |Q|$. Moreover, $R_{AdS_2}^2$ does not scale with $Q$, so we have inverse scale separation as for the large BH example in toy model 2. The intuition is that this branch of solution represents black holes pulling the scalars towards a region where the scalar potential $V$ is not flat, therefore keeping the scalars close to their cosmological minimum.

\paragraph{The right branch:}\ \\
As we will show below, we find on the right branch another behavior. As one increases $Q$, the value of $x$ on the right branch reaches the right edge of the allowed range of $x$ - there, $e^{2D}\to 0$ and $\nu_3 \to \infty$. More precisely, if one sets 
\begin{align}
    x = x_R -\epsilon\, ,
\end{align}
we must scale $\tilde{Q} \propto \epsilon^{-1}$. This in turn gives large $S^2$ radii in AdS units:
\begin{align}
    R^2 &\propto \epsilon^{-3/2} L_{AdS_4}^2 \, .
\end{align}
With this choice, one finds that the scalars $\nu_a$ ($a=1,2,4$) are proportional to their vacuum value with an $\mathcal{O}(1)$ proportionality constant, whereas $\nu_3$ and $D$ scale with $\epsilon$ as:
\begin{align}
    \nu_3 &\propto \epsilon^{-1/2} \nu_3^{vac} \, , \quad e^{D} \propto \epsilon^{1/2} e^{D^{vac}} \, ,
\end{align}
where the precise proportionality constants are unimportant to our discussion. The above shows that we have solutions with arbitrarily large scalar distances for large BHs. More precisely, since the moduli space metric in the $(\nu_3,D)$ directions is flat in the variables $(\ln\nu_3,D)$, $\dd s^2=\frac12(\dd\ln\nu_3)^2+2\,\dd D^2$ (from \eqref{eq:real_kinetic_matrix}), the geodesic distance from the vacuum grows as
\begin{equation}
\Delta= \frac{\sqrt{10}}{3}\ln\frac{R}{L_{AdS_4}}+\mathcal{O}(1)\,,
\end{equation}
i.e.\ logarithmically in the size of the black hole in AdS units. This is precisely the behaviour found for the string-like Model~3 in the weak-coupling direction (large BHs with $e^{a\phi}\to 0$). It is therefore natural to ask whether the DGKT potential vanishes in this asymptotic region. To do so, note that on our ansatz one has $e^{2D}\nu_3^2 = \frac{2x^2}{N^2}$, leading to $\nu_3^2 \propto e^{-2D}$ on the edge of the right branch where $x\to x_R$. Plugging this relation in the potential, along with the values of the $\nu_a$ scalars, one finds
\begin{align}
    V &= -\frac{1+\sqrt{13}}{2\sqrt{2}(3+\sqrt{13})}e^{3D} + \frac{\sqrt{7+2\sqrt{13}}}{6\sqrt{2}(3+\sqrt{13})}(\alpha^2-2)N^3 e^{5D}
\end{align}
which indeed approaches 0 from below as $D\to-\infty$. We indeed have the same behavior as in toy model 3, where large BHs can indeed displace scalars arbitrarily far from the vacuum if this happens in a flat direction of the potential. Another similarity with toy model 3 is the behavior of the squared AdS$_2$ radius: here it is equal to $\frac{1}{2}R^2$ at leading order in $\epsilon$, i.e.\ $R^2V(\phi_h)\to-\frac12$ by \eqref{eq:RAdS2_eq_AdS2S2_top_down}: as in Model 3, the horizon size stays comparable to the local scale set by the potential at the horizon value of the scalars. There is therefore no 2d scale separation in the large $S^2$ radius limit of the ``right branch'', which is the same result as that of the fixed $Q$, large $P$ limit of toy model 3.\\

It is also important to understand whether this $\epsilon \to 0$ limit is under control since $\nu_3$ blows up. A proxy for this is whether the string coupling remains small. In fact, the blowing up of $\nu_3$ is balanced by the suppression of $e^{D}$ compared to the vacuum, leading to the following behavior of $g_s$ at small $\epsilon$:
\begin{align}
    g_s^2 &= \frac{2}{3} \sqrt{\frac{2}{5} \left(13+\sqrt{13}\right)} (g_s^{vac})^2 \sqrt{\epsilon}\, ,
\end{align}    
where $g_s^2=e^{2D}\mathcal{V}_6$. The internal volume $\mathcal{V}_6\propto\nu_3$ grows as $\epsilon^{-1/2}$, so the branch goes to weak coupling and large volume, where the supergravity approximation is under control. Since only $\nu_3$ (and $e^{-D}$) blows up while the other moduli stay at $\mathcal{O}(1)$ multiples of their vacuum values, this is presumably a (partial) decompactification limit, in which, in line with the distance conjecture, a tower of KK states becomes light. We do not analyse this limit further here.

\paragraph{AdS$_2$ stability}\ \\
One could then wonder if these BR solutions on either branch are stable. As explained in the appendix, the spectrum of masses around AdS$_2$ is given by the eigenvalues of the matrix \eqref{eq:AdS2_mass_matrix}. For our purposes, the computation of the spectrum goes as follows: for any charge $Q$, find the value(s) of $x$ numerically by solving \eqref{eq:Fx_eq}. Then, evaluate the mass matrix for this value of $x$ and find the eigenvalues.\\

However, there is something to be understood analytically as well: it turns out that the mass matrix in our case exhibits a block triangular structure, where one block is 1x1. This block is in the $U$ direction. In other words, the equation for the fluctuation $\delta_U$ is decoupled from the other ODEs. Indeed,
\begin{align}
    \square_2 \delta^U
    &=  2U_*^2 \left(\partial_n \partial_U V\right)_{\phi^k=\phi^k_*} \delta^n\, ,
\end{align}
where we used $G^{Uk}=2U^2 \delta^{Uk}$ and the fact that $V_{BH}$ is $U$-independent. It turns out that when evaluated on the AdS$_2$ vacuum, the gradient of $\partial_U V$ only has nonzero components in the $U$ direction, therefore showing the ODE decoupling from the other directions. When explicitly plugging in our ansatz, we get
\begin{align}
    m_U^2 &= \frac{2}{N^{9/2}(\alpha^2-2)^{3/2}}  \frac{\left(3\sqrt{x^2+4} - x^2 - 3\right)^{3/2} x^{3/2}}{(x^2+4)^{5/4}}  \left(2 - (4 + x^2)^{1/2}\right)
\end{align}
Now we can compute $m^2_U R_{AdS_2}^2$ and check whether that is above or below the BF bound of $-\frac{1}{4}$. Using our ansatz we get the remarkably simple expression
\begin{align}
    m^2_U R_{AdS_2}^2 = -\frac{y-2}{y^2-y-3}\,,\qquad y=\sqrt{x^2+4}\,.
\end{align}
In the range $x_c<x<x_R$ corresponding to the two branches, this function increases monotonically from $-1+\frac{1}{\sqrt3}\simeq-0.423$ at $x_c$ to its value at the right edge:
\begin{align}
    m^2_U R_{AdS_2}^2 \leq \frac{\sqrt{13}-7}{12}\simeq-0.283 < -\frac{1}{4}\, ,
\end{align}
and therefore the $U$ mode is below the BF bound in 2d. Hence \emph{all} the Bertotti-Robinson solutions within this truncation, on both branches, are perturbatively unstable. The instability is in the complex structure modulus $U$, which is not sourced by the black hole. Whether stable solutions exist outside the truncation (e.g.\ with $P\neq0$ or less symmetric fluxes), and what the endpoint of the instability is, are important open questions. The physics described above would have to be reconsidered in light of the full BH solution, when it is known which of the two branches connects to the vacuum at infinity through a full black hole solution.

\section{Discussion}\label{sec:discussion}

We studied how charged black holes displace scalar fields towards their horizon. Such a displacement is interesting because large field displacements lead to light towers of particles \cite{Ooguri:2006in}, allowing UV physics to become relevant at horizon scales.  However, cosmological moduli stabilisation might be an obstacle to realize significant field displacements. Then, only black holes with radii probing the UV scale can displace the scalars of a compactification \cite{Green:2006nv}. We demonstrated this explicitly in our first toy model (section~\ref{sec:model1}), a massive scalar around a Minkowski minimum: for $mR\gg1$ the displacement vanishes as $1/(mR)^2$, so only black holes smaller than $m^{-1}$ displace the scalars. A Minkowski vacuum is in this regard similar to an (A)dS with a ``small" cosmological constant. For that reason we dubbed compactifications where black holes behave as in this toy model to be \emph{rigid}: the KK scale is scale-separated from the (A)dS scale \textbf{and} the masses of the moduli should be parametrically higher than the (A)dS scale.
Since larger field displacements come together with UV degrees of freedom becoming light we conclude that physics from the UV completion (the tower) becomes relevant at horizon scales only for small enough black holes in rigid compactifications. In this respect the fuzzball proposal is in tension with having rigid compactifications, since it requires large field displacements (see Appendix \ref{App:FUZZ}).

The picture changes when the scalar mass is tied to the cosmological constant, as in Models 2 and 3 and in all known scale-separated vacua. Black holes are then ``small'' when they are small compared to the (A)dS radius, which for our universe is the Hubble length. Such black holes drag the scalar over large distances, while larger ones still displace it by an $\mathcal{O}(1)$ amount, because their horizon carries a finite electromagnetic energy density. With Hubble-scale moduli masses, astrophysical black holes are small in this sense: Sagittarius A$^*$ could displace a modulus by $\Delta\approx52$ Planck units in Model 2. In the string-like Model 3, large black holes can drag the scalar arbitrarily far along the direction where the potential vanishes, for suitable charges. 

Indeed, rigid string compactifications are not easy to come by and might simply not exist. To clarify the situation we constructed what are, to our knowledge, the first charged near-horizon solutions in a top-down scale-separated flux vacuum (DGKT on $\mathbb{T}^6/\mathbb{Z}_4$), where the moduli masses are order one in AdS units. Within our truncation these solutions exist only above a minimal charge, i.e.\ for horizons larger than the AdS radius. They come in two branches: on one the moduli are displaced by an $\mathcal{O}(1)$ amount (as in Model 2), on the other they run off to weak coupling and large volume with a distance growing logarithmically with the horizon size (as in Model 3). All of them are perturbatively unstable, and whether they are the endpoint of full black hole flows remains open. We expect sizable displacements for black holes that are large in Planck units to be a generic feature of non-rigid vacua. One could wonder whether the picture would change if we were able to engineer a detailed embedding of our toy model in a Type IIB Minkowski no-scale flux vacuum, as sketched in Appendix \ref{App:IIB}. However, as discussed in section \ref{sec:topdown}, once K\"ahler moduli stabilization is taken into account, the black hole solution will be altered and will likely source a profile for the K\"ahler moduli along the complex structure moduli. The resulting compactifications are not rigid (due to the light mass of the K\"ahler moduli \cite{Gautason:2018gln}), and so we expect to recover the same qualitative results as in the IIA AdS case.\footnote{It would be interesting to understand how the picture might differ in the recently proposed rigid vacua of \cite{Rajaguru:2024emw,Becker:2024ayh}.}

We now want to argue for Swampland principles based on these observations.  If we insist that for \emph{all black hole sizes} at least some UV physics become relevant at horizon scales, we need to conclude that this can only occur in vacua that are not rigid and hence we speculate that
\begin{center}
\underline{Swampland Speculation}: \emph{Rigid compactifications are in the Swampland.}\\
\end{center}
We are aware of at least two Swampland conjectures that challenge rigidity and hence support our Speculation; 1) The strong version of the AdS distance conjecture \cite{Lust:2019zwm} excludes scale separation for SUSY AdS vacua.  2) The AdS moduli conjecture \cite{Gautason:2018gln}(see also \cite{Blumenhagen:2019vgj}) entails that the lightest scalar is not parametrically heavy in AdS units. This is true for all known AdS vacua with claimed scale separation, ever constructed. It poses a major challenge for string phenomenology because, when applied to our universe, we would have a scalar that is 30 orders lighter than the neutrino mass scale.

Reasons why UV physics has to play a role at horizon scales, and hence puts rigid compactifications in the Swampland, could be any of the following; 1) To accommodate for black hole microstates we must have corrections to black holes at horizon scale. 2) To allow unitary black hole evaporation we want sufficient deviations from effective field theory, be it non-locality, or something else.  There is no consensus as to whether UV physics is needed at horizon scales to accommodate microstates or unitarity, see for instance \cite{Mathur:2009hf} and \cite{Raju:2020smc} for competing views.

One could worry that our conclusions are based on thought experiments with charged black holes and so have no bearing on realistic neutral black holes. We believe this worry is not justified. The main reason for focusing on charged extremal solutions was computational simplicity; using the attractor mechanism one can readily compute the positions of the scalar fields at the horizon.  However, the charge is not crucial \cite{Green:2006nv}; if we take a string theory perspective on our own observed universe we have to conclude that the Standard Model parameters (fine structure constant, Yukawa couplings,...) are vevs of scalar fields.\footnote{Although this can lead to semantic tensions \cite{Banks:2025nfe,Sen:2025bmj}.}  Symbolically we have:
\begin{equation}\label{eq:SM}
\frac{2\kappa_4^2\mathcal{L}}{\sqrt{-g}} = 
\frac{\mathcal{R}}{2} - \frac{1}{2}(\partial\phi)^2- \frac{1}{4}f(\phi)\text{Tr}F^2 - i\bar{\psi}\slashed{D}\psi - g(\phi)\bar{\psi}\psi - V(\phi)\,,
\end{equation}
where all field indices on scalars, fermions and vectors have been suppressed. Hence, aside the Higgs (which is part of the symbol $\phi$ in \eqref{eq:SM}), all stringy moduli will also couple to the fermions. Secondly, it appears that the black holes observed in our universe originate from collapsing matter clouds made up of fermion condensates, known as stars. Hence, the expectation is that stars displace moduli very mildly, but as they collapse towards reaching critical density, the displacement will grow. When it eventually rings down to a Schwarzschild (Kerr) solution one usually truncates all the matter and considers only the Einstein-Hilbert action. Therefore, one could be tempted to conclude the displacement of the scalars vanishes. This, in our opinion, seems unlikely because we have no full understanding what is inside a black hole nor can we claim the fermion condensates ceased to exist. Yet, the no hair theorem seems to suggest naively the displacement is absent. We do not take this point of view, like \cite{Green:2006nv}, and consider, as a safe bet, that the displacement is at least as large as the displacement right before reaching critical density. It is natural to expect the scalar field displacement to be subject to an effective black hole potential controlled by
\begin{equation}
V_{BH}(\phi)\sim  g(\phi)\langle \bar{\psi}\psi\rangle\,.   
\end{equation}
It would be very interesting to make this all much more precise in concrete models.\footnote{It could suffice to consider the Standard Model and TeV-sized black holes that displace the Higgs field. The Higgs potential has been argued to have an AdS stable region \cite{Elias-Miro:2011sqh} and one can wonder whether the Higgs can be pushed all the way into that region as to trigger vacuum decay in our universe. In the special case the AdS bubble created this way could stabilize one would recover the black hole mimicker mechanism suggested by Danielsson et al \cite{Danielsson:2017riq}. In more realistic scenarios such that of \cite{Maldacena:2020skw,Gervalle:2024yxj} magnetically charged black holes with electroweak hair were considered. If the magnetic charge is tuned the right way, small black holes can push the Higgs off of its minimum and form a bubble of false vacuum at the horizon. Such black holes can have a rich phenomenology, as seen in e.g. \cite{Bai:2020spd,Bai:2021ewf,Ghosh:2020tdu,Estes:2022buj,Diamond:2021scl}.} For now we just take as a principle that displacement of scalar fields is not specific to charged black holes but should apply certainly as well to more physical black holes that result from collapsing matter. Hence the qualitative behavior of scalar displacements is not tied to charged black holes.

Finally, we want to iterate an important observation made in \cite{Green:2006nv}. If instead of SUSY AdS vacua we would consider black holes in meta-stable vacua there is the intriguing possibility that black holes push moduli into the true vacuum and cause a bubble of true vacuum to form that overtakes the entire universe.  The fact that our universe has not decayed yet into some SUSY AdS vacuum, despite the presence of black holes, can lead to constraints on the possible consistent compactifications. It would be interesting to study this more in-depth.

\section*{Acknowledgments}
We have benefited much from discussions with Daniel Mayerson on fuzzball geometries and thank him for writing Appendix \ref{App:FUZZ}. We furthermore would like to thank Steven Abel, Iosif Bena, Naomi Gendler, Jacob Moritz, Erik Plauschinn, Angel Uranga and Cumrun Vafa for useful discussions.\footnote{We do not thank Miguel Montero.} S.R. is supported by the Research Foundation - Flanders (FWO) doctoral fellowship 11PAA24N, and also acknowledges support from the FWO Odysseus grant G0F9516N and the KU Leuven iBOF-21-084 grant.

\appendix

\section{Black hole flow solution for model 1}
\label{app:BH_model_1}

We want to find large black hole solutions of the theory \eqref{eq:generic_gravity_action} with $V=\frac{1}{2}m^2\phi^2$ by using the 1d effective action formalism \eqref{eq:S_eff_generic_fields}. To find large black hole solutions perturbatively, we use the small parameter 
\begin{equation}
\epsilon=\frac{1}{(mR)^2}\,.    
\end{equation}
At 0th order, the dilaton is completely frozen at $\phi=0$ and our theory effectively reduces to pure Einstein-Maxwell theory and the unperturbed solution is the Reissner-Nordstrom (RN) black hole
\begin{align}
    \dd s^2 &= -e^{2U(z)}\dd t^2 + e^{-2U(z)}\left(\frac{c^4}{\sinh^4(cz)} \dd z^2 + \frac{c^2}{\sinh^2(cz)} \dd \Omega^2\right) \, , \\
    F &= P \dd \Omega + Q e^{2U(z)} \dd t \wedge \dd z\,\\
    e^{-U(z)} &= \cosh(cz)-\frac{M}{c} \sinh(cz), \quad c=\sqrt{M^2-\frac{1}{4}(P^2+Q^2)}\, ,
\end{align}
where $c$ is the extremality parameter. For computational simplicity, we focus on extremal large black holes, so we set $c$ to 0 from now on (and therefore $M^2=\frac{1}{4}(P^2+Q^2)=R_0^2$). 

The perturbation theory ansatz is then given by
\begin{align}
    U(z) &= \sum_{k=0}^{\infty} U_k(z)\epsilon^k, \quad e^{-U_0(z)} = 1-Mz\, ,\\
    A(z) &= \sum_{k=0}^{\infty} A_k(z)\epsilon^k, \quad 
    e^{2A_0(z)} = \frac{1}{z^2}\, ,\\
    \phi(z) &= \sum_{k=1}^{\infty} \phi_k(z)\epsilon^{k}\, .
\end{align}
and we have to solve the following equations of motion (deriving from \eqref{eq:S_eff_generic_fields}) 
\begin{align}
    \ddot{\phi} &= e^{2U}V_{BH}'(\phi) + V'(\phi) \, e^{-2U+4A} \, ,\\
    \ddot{U} &= 
    - \frac{1}{2}e^{-2U+4A}V(\phi)
    + \frac{1}{2}e^{2U}V_{BH}(\phi)\, \\
    \ddot{A} &= 
    e^{2A}-V(\phi)e^{-2U+4A}\, ,\\
    2\dot{U}^2 - 2\dot{A}^2 + \frac{1}{2}\dot{\phi}^2 &= e^{2U}V_{BH}(\phi)
    + V(\phi)e^{-2U+4A}
    - 2e^{2A}\, ,
\end{align}
perturbatively in $\epsilon$. Note that the series for $\phi(z)$ starts at $k=1$ so that in the ``large $m$" limit ($\epsilon \to 0$) the dilaton is stabilized at 0, as it should. Said differently, this is required for consistency of the perturbative analysis of the above EOMs. The dilaton equations of motion at leading order in $\epsilon$ is therefore just an algebraic equation and leads to
\begin{align}
    \phi_1(z) &= -\frac{a}{2}R_0^2e^{4U_0-4A_0}\left(P^2-Q^2\right)\, .
\end{align}
whose horizon value and interpretation are discussed in section~\ref{sec:LBHtoymodel}.

We now turn to describing how the geometry is affected by the presence of the scalar. At leading order, the correction to the metric is given in terms of the functions:
\begin{align}
    A_1 &= C_A (R_0z)^2 + \frac{D_A}{R_0 z} + \frac{a^2 \left(P^2-Q^2\right)^2}{16 R_0^4}\left(
    4\frac{\ln(1-R_0z)}{R_0z}
    +\frac{P_A(1-R_0z)}{R_0z(1-R_0z)^4}
    \right)\, , \label{eq:A1_RN_general}\\
    U_1 &= C_U (1-R_0z)^2 + \frac{D_U}{1-R_0z} + \frac{a^2\left(P^2-Q^2\right)^2}{16 R_0^4}\left(
    -4 \frac{\ln(1-R_0z)}{1-R_0z} + \frac{P_U(1-R_0z)}{(1-R_0z)^4}\right)\, , \label{eq:U1_RN_general}
\end{align}
where $C_A(R_0z)^2+D_A/(R_0z)$ and $C_U(1-R_0z)^2+D_U/(1-R_0z)$ are the general solutions of the homogeneous equations $\ddot A_1=2e^{2A_0}A_1$ and $\ddot U_1=2R_0^2e^{2U_0}U_1$, and $C_A$, $D_A$, $C_U$ and $D_U$ are integration constants which we fix below and $P_U, P_A$ are the following polynomials:
\begin{align}
&P_U(x) = \frac{3}{2}x^4 
    -\frac{9}{2}x^2+\frac{6}{5}x-\frac{1}{6}\, ,\\
&P_A(x) =-x^5 + \frac{13}{2}x^3 -\frac{27}{10}x^2 +\frac{23}{30}x -\frac{1}{10}\,.
\end{align}
It is straightforward to plug our solutions for $A_1$ and $U_1$ into the constraint equations and see that it reduces to $C_A=C_U\equiv C\, $. Imposing the fact that we want flat space at infinity in the usual coordinates further fixes:
\begin{align}
    D_A &= - \frac{13}{60} \frac{a^2 \left(P^2-Q^2\right)^2}{R_0^4}\, , \quad
    D_U = -C + \frac{59}{480}\frac{a^2\left(P^2-Q^2\right)^2}{R_0^4}\,.
\end{align}
Therefore, all that remains is the constant $C$. It can be fixed by imposing extremality along the $\epsilon$ deformation. To do so, we require the presence of an AdS$_2\times S^2$ near-horizon geometry throughout the deformation, which requires $C=0$ (the mode $(1-R_0z)^2$ grows at the horizon $z\to-\infty$ and would destroy the AdS$_2$ throat). With $C=0$ one obtains \eqref{eq:U1_RN}--\eqref{eq:A1_RN} and the near-horizon geometry \eqref{eq:AdS2xS2_R_AdS}--\eqref{eq:AdS2xS2_phi} of the main text.

\section{Toward Black Holes in IIB Flux Compactifications} \label{App:IIB}

We aim at a top-down construction of the toy model of section \ref{sect:bottom_up_EMD_theory} into IIB string theory. Specifically, we consider Type IIB on a Calabi-Yau (CY) orientifold with appropriate fluxes, yielding a 4d $\mathcal{N}=1$ no-scale Minkowski vacuum where all relevant complex structure moduli are stabilized and the K\"ahler moduli are left unstabilised \cite{Giddings:2001yu}. This setup includes $U(1)$ gauge fields whose gauge kinetic functions depend on the stabilized moduli, allowing for the construction of charged black holes that compete with the stabilizing potential.  In what follows we ignore the effects of K\"ahler moduli stabilisation and come back to it at the end.

This type of scenario was examined in \cite{Danielsson:2006jg}, where the effect of a charged black hole on stabilized complex structure moduli was studied. They pick a stabilization scheme where the complex structure moduli are stabilized near the (deformed) conifold point. Such a choice is phenomenologically relevant and constitute the first step of the KKLT \cite{Kachru:2003aw} construction. The fluxes are chosen such that scalars are stabilized at a mass typically close to the string scale\footnote{Unless one considers the moduli associated to local cycles down the throat} and the potential leads to a Minkowski vacuum, at least at the classical two-derivative level. Their analysis, valid in the regime $mR \gg 1$, qualitatively agrees with section \ref{sec:LBHtoymodel}: large black holes do not destabilize the vacuum. They also comment on the fact that the picture changes when one instead considers small black holes or very light moduli, consistent with section \ref{sec:SBHtoymodel}.

While the construction in \cite{Danielsson:2006jg} captures the essential physics, several subtleties remain before it can be considered a complete string-theoretic embedding. In particular, \cite{Danielsson:2006jg} treats the black hole as a solution of the parent 4d $\mathcal{N}=2$ theory, where it can be realized as D3-branes wrapping 3-cycles in the CY. However, in a CY orientifold with fluxes, many such cycles are projected out or carry fluxes, making the wrapping of D3-branes a lot more subtle. Indeed, although the construction of charged black holes in flat 4d $\mathcal{N}=2$ CY compactifications of type IIB is well studied \cite{Denef:2000nb}, once we orientifold and add in fluxes, very little is known about how black holes can be constructed. The remainder of this section is devoted to clarifying these subtleties, paving the way to study black holes in  a fluxed CY orientifold in full detail.

We first recall some elements of CY orientifold compactifications in type IIB (and point the reader to \cite{Grana:2005jc,Grimm:2005fa} for a detailed discussion). Under the involution $\sigma$ of the orientifold, the cohomology of the parent CY manifold splits into even and odd eigenspaces:
\begin{equation}
    H^p(Y) = H_{+} ^p \oplus H _ { -} ^p\,.
\end{equation}
In particular, this splitting occurs for the harmonic forms that give the vector multiplets $\{z^i, A^i\}$ in 4d and the orientifold acts in such a way that even (resp. odd) scalars and odd (resp. even) gauge fields are projected out. To be concrete, let us consider a Type IIB O3/O7 orientifold which keeps only the complex structure moduli $z ^k $ and the vectors $A^\alpha _ \mu$, with $k= 1,..., h_{2,1}^{-}$ and $\alpha= 1,..., h_{2,1}^{+}$. All other scalars $z^\kappa$, vectors $A_\mu^k$ and the graviphoton are projected out of the theory. We will use Greek letters for even components and Latin letters for odd components. Whether or not a 4d field is projected out is determined by the properties of the 10d field from which it stems. Indeed, the 10d fields are all either odd or even under the involution and so, under dimensional reduction over the CY, their components get decomposed onto bases of $H_{+} ^p $ and $ H _ { -} ^p$ in the CY. In particular, the middle cohomology $H^{(3)}$ splits into $H^{(3)}_{+} \oplus H^{(3)}_{-} $: its real symplectic basis splits into $(\alpha_{\kappa},\beta^{\lambda})$ for $H^{(3)}_{+}$ and $(\alpha_{\hat k},\beta^{\hat l})$ for $H^{(3)}_{-}$. The holomorphic three-form $\Omega$ is odd under the involution, so that only its $H^{(3)}_{-}$ part survives: 
\begin{equation}
    \Omega_3 ( z^{k}) = Z^{\hat k} \alpha _ {\hat k} - \mathbf{F}_{\hat k}\beta^{\hat k}\,.
\end{equation}
where the hatted coordinates go from $0$ to $h^{(2,1)}_{-}$, whilst the un-hatted ones span $\{1,..,h^{(2,1)}_{-}\}$. As is familiar from the $\mathcal{N}=2$ case, there are $h^{(2,1)}_{-}$ physical complex structure moduli that can be taken to be the projective coordinates $z^k= Z^{\hat k}/Z^0$. On the other hand, the gauge fields $A^\alpha _\mu$ come from the ten dimensional RR 4-form, which is even under the involution: it leads to $\frac12 h^{(3)}_{+}= h^{(2,1)}_{+}$ vectors $A^{\kappa}_\mu$. Importantly, the gauge kinetic functions of these U(1) vector fields are simply those of the parent 4d $\mathcal{N}=2$ theory on which one imposes that all even moduli $z^\kappa$ vanish. This means that we obtain an EFT in 4d with $h^{(2,1)}_{+}$ U(1) vectors $A^\alpha _ \mu$, whose kinetic terms only couple to the $h^{(2,1)}_{-}$ $z^k$ that are left over after the orientifold involution. This points to the fact that one could potentially wrap D3 branes on even cycles in the CY and obtain a black hole that acts on the odd complex structure moduli $z^k$. However, we should not forget about the fluxes, which source the stabilizing potential for the scalars and which can lead to topological restrictions for wrapping branes on cycles. Indeed, stablization is obtained by generating flux backgrounds for the NSNS and RR 3-form field strengths in the CY however, wrapping D3 branes on cycles with non-trivial NSNS (resp. RR) 3-form flux, generates Freed-Witten anomalies which lead to the D3 emitting D1 \cite{Maldacena:2001xj} (resp. F1 \cite{Witten:1998xy}) strings. To avoid such complications, one should therefore hope that the background 3-form fluxes are cohomologically trivial on the world-volume of the D3 branes that make up the black holes. Luckily, it turns out that this is the case: both $F_3$ and $H_3$ are odd under the orientifold involution and are therefore parametrized by elements of $H_{-}^{3}(Y)$.

It therefore seems that we have all the ingredients necessary to build black holes that source a potential for the $z^k$ which are stabilized by background fluxes. A few comments are in order.  In \cite{EnriquezRojo:2020hzi}, it is explained how placing a D3 on the even cycles breaks all of the supersymmetries. Therefore, any black hole built in this way will most likely be unstable. Whilst this is not necessarily of much importance for our purposes, it partly explains why so little is known about black holes in 4d $\mathcal{N}=1$ compactifications of string theory. Secondly, note that in the above discussion, we have completely ignored the moduli coming from the hypermultiplets of the parent 4d $\mathcal{N}=2$ theory. The reason is that they do not enter in the gauge kinetic functions and they decouple from the complex structure moduli in the metric on moduli space in the large complex structure limit. Once fluxes are introduced, they generate a potential that allows one to stabilize (some of) the complex structure moduli and the dilaton. Another effect that we have not discussed here is the presence of spacetime filling D3 or D7 branes that are introduced to cancel the RR-tadpole induced by the fluxes. We are implicitly assuming that the open string sector completely decouple from the closed string physics discussed in this section. Finally, it was  argued in \cite{EnriquezRojo:2020hzi} that wrapping D3 branes on these even cycles leads to massless particles (and not black holes). The reason is that the mass of such particles is usually given by the volume of calibrated (BPS) cycles is by integrating the holomorphic 3-form over them. This, by definition, vanishes for even cycles and for that reason \cite{EnriquezRojo:2020hzi} states that instead of black holes, one obtains massless particles in 4d. We believe this is misleading since the even cycles are non-BPS and this is not the way to measure their volume.\footnote{We thank Jacob Moritz for helpful discussions about this.} Properly estimating the volume of these non-BPS cycles could lead to non-BPS black holes as the ones discussed in \cite{Long:2021lon}.

All in all, barring the subtleties we just mentioned, it should at least be possible to study how a black hole solution charged under the $h^{(3)}_{+}$ U(1)'s should affect the stabilized complex structure moduli from the perspective of the EFT. Finding an explicit embedding comes down to finding a CY orientifold that allows for fluxes that stabilize the complex structure moduli that couple to the U(1)'s under which the black hole is charged. Then, one would have to solve the attractor equations. The main difference between such a model and the toy model in section \ref{sect:bottom_up_EMD_theory}, except for the added complexity for having more moduli and more gauge fields, is the fact that the scalar potential will not be a simple $\phi^2$ potential. Nevertheless, in the large complex structure limit of such string theoretic models, one can generally expect (see e.g. \cite{Ooguri:2018wrx,VanRiet:2023pnx}) the scalar potentials to be given in terms of polynomials in exponentials of the canonically normalised saxions. Therefore, around the minimum of the potential, and for $mR \gg 1$, we expect to recover similar physics as in the toy model, as is confirmed by \cite{Danielsson:2006jg}. In the opposite limit $mR \ll 1$ where the stabilization potential is negligible, we expect once more to recover the standard attractor black holes which can always drag the moduli to large distances in moduli space.

In this section we have studied the embedding of our toy model in IIB string theory in a context that lacks full moduli stabilisation since it relies on no-scale Minkowski backgrounds. Those backgrounds will certainly be altered after also K\"ahler moduli stabilisation is taken into account. Whether or not the black hole solutions will be strongly altered by this depends on whether the K\"ahler moduli couple to the vectors after full moduli stabilisation. Famous suggestions on how to achieve full moduli stabilisation in this context are based on the so-called KKLT scenario \cite{Kachru:2003aw} (or its modern follow-ups \cite{Demirtas:2021nlu}) or the Large Volume Scenario \cite{Balasubramanian:2005zx}. Unfortunately these scenarios lack a full top-down understanding and as a consequence their consistency is still being debated, see for instance \cite{Lust:2022lfc, Bobev:2023dwx}.

\section{The $\mathbb{T}^6/\mathbb{Z}_4$ orbifold: the details}
\label{app:T6_Z4_orbifold_details}
We now discuss the aspects of the $\mathbb{T}^6/\mathbb{Z}_4$ model that are relevant for the computations in section \ref{sec:topdown}, following \cite{Ihl:2006pp, Ihl:2007ah}. First, let us summarize the various untwisted moduli of this model. First, there are the complex K\"ahler moduli $t_ a= u_ a + i \nu _a$ coming from decomposing the complexified K\"ahler form ($iJ+B$) on the basis of $h^{1,1}_{-}=4$ odd $(1,1)$-forms. Then, the RR 3-form and the holomorphic 3-form also yield 2 sets of moduli when decomposed on the basis of 3-cycles ($a_K$ are even and $ b_K$ are odd):
\begin{equation}
   C_3 = \xi_ K a _ K\,,\qquad\;\;\; \Omega= \mathcal{Z}_K a_ K - \mathcal{F}_ K b_ K\,.
\end{equation}
where the $\xi_K$ and $Z_K$ are both real and $\mathcal{F}_K$ is pure imaginary. We have the following constraint on the holomorphic three-form: \begin{equation}\label{eq:cons}
    Z_ K \mathcal{F} _ K = - i /2\,.
\end{equation}
There is only one untwisted complex structure modulus, denoted by $U$, which is a real variable that takes values in $]0, \infty[$. The $Z_K$ and $\mathcal{F}_K$ depend on it as follows: 
\begin{align}\label{eq:z1z2}
    Z_1 &= \frac{1}{2 \sqrt{U}}\left(\frac12 + U\right)  & Z_ 2 & = \frac{1}{2 \sqrt{U}}\left(\frac12 - U\right) \\ \mathcal{F}_1 &= - i Z_1 &  \mathcal{F}_2 &= i Z_2
\end{align}
For convenience, the RR 3-form $C_3$ can be combined with the holomorphic 3-form $\Omega$ and the dilaton to give the complexified holomorphic three-form:
\begin{equation} \label{eq:Omegac}
    \Omega_ c = C_3 + 2 i e^{-D} \text{Re} [\Omega]  = ( \xi _ K + 2 i e^{ - D} Z_ K ) a_ K  = 2 N_K a_ K
\end{equation}
where the four dimensional dilaton $D$ is linked to the ten-dimensional one through $e^{-D}= \mathcal{V}_6^{1/2} e^{-\phi}$, and the volume is given by $\mathcal{V}_6 = \frac{1}{6}\kappa_{abc} \nu _ a \nu_b \nu_ c=\frac14 \nu_3 ( \nu _1 \nu_2 - 2 \nu_4^2)$. Decomposed on the basis of two independent even 3-cycles, $\Omega_c$ yields two moduli $N_K$, with $K \in \{1,2\}$.

In 4d $\mathcal{N}=1$ language, the effective action is given by:
\begin{equation}
    S^{(4)} = \int  \left(\frac12 R \star 1  - K_{ I \bar J } \star d \phi ^I \wedge d\bar{\phi} ^{\bar J} - V \star 1 - \frac12 (\text{Re} f_{\alpha \beta}) \star F^\alpha \wedge  F^\beta - \frac12 ( \text{Im} f_{\alpha \beta}) F^\alpha \wedge F ^\beta \right)
\end{equation}
where the scalar potential is given by 
\begin{equation}\label{eq:scalpot_app}
    V= e^K \left(K^{I\bar J} D_ I W D_{\bar J } \bar{W} - 3 |W|^2\right) + \frac12 (\text{Re} f) ^{- 1 \alpha \beta}D_ \alpha D_ \beta\,,
\end{equation}
and where $\star$ is the four-dimensional Hodge star, $K_ { I \bar J } = \partial _ I \partial _ { \bar J } K$, and $D_ I W = \partial _ I W + ( \partial_I K) W$ is the K\"ahler covariant derivative. For the $\mathbb{T}^6/\mathbb{Z}_4$ orbifold, one can show that the D-term contributions to the scalar potential vanish and that the K\"ahler potential is given by: 
\begin{equation}\label{eq:kalpot}
    K= 4 D - \log ( 8 \mathcal{V}_6)\,.
\end{equation}
From this, we obtain the inverse metric on moduli space:
\begin{equation}
    K^{I \bar J} =  \begin{pmatrix} e^{ -2 D}(Z_1^2 + Z_2^2) & 2 e^{ -2 D} Z_1 Z_2 & 0&0&0&0 \\  2 e^{ -2 D} Z_1 Z_2 & e^{ -2 D}(Z_1^2 + Z_2^2)& 0&0&0&0\\ 0&0& 4 \nu_1^2 & 8 \nu_4^2 & 0 & 4 \nu_1 \nu_4\\
    0&0&8 \nu_4^2 & 4 \nu_2^2 & 0 & 4 \nu_2 \nu_4 \\ 0&0&0&0&4 \nu_3^2 & 0 \\0&0& 4 \nu_1 \nu_4 & 4 \nu_2 \nu_4 &0& \nu_1 \nu_2 + 2\nu_4 ^2
        \end{pmatrix}^{I \bar J} \,, \label{eq:inverse_metric_moduli_space}
\end{equation}
(in the basis $(N_1,N_2,t_1,t_2,t_3,t_4)$), where we used \eqref{eq:cons} and \eqref{eq:Omegac} to write:
\begin{equation}
      2 + e^{2 D} \left( ( N_1 - \bar{N_1})^2 - ( N_2 - \bar{N_2})^2\right) = 0\,,
\end{equation}
which yields the following derivatives of $D$, which enter the K\"ahler covariant derivatives through \eqref{eq:kalpot}:
\begin{align}
    \partial_ { K} D &= \frac{\partial D}{\partial N _ K} = - \mathcal{F}_ K e ^D\,,\\
    \partial_ {\bar K} D &= \frac{\partial D}{\partial \bar{N} _ K} =  \mathcal{F}_ K e ^D\,.
\end{align}

\subsection{The Flux Contribution}
\label{app:flux_contribution}
We now turn to the allowed fluxes which we can turn on in this model. The presence of these fluxes will generate a non-vanishing superpotential $W$ which will have the effect of creating a stabilizing potential for the K\"ahler moduli and dilaton. Importantly, the addition of these fluxes has no effect on the kinetic terms of the four-dimensional fields nor does it change the K\"ahler potential. We stick to the classification of RR fluxes in terms of cohomology and expand the RR fluxes on the appropriate basis of even and odd forms:
\begin{equation}
    F_0 = m_0,\;\;F_2 = m_a \omega_a, \;\; F_4 = e_ a \tilde{\omega} _ a,\;\; F_6 = e_0 \varphi\,,
\end{equation}
where $\varphi$ is the odd 6-form, $\omega_a$ are the 4 odd (1,1)-forms (the same basis in which $J$ and $B$ are expanded to define $t_a$) and $\tilde{\omega}_a$ are the four dual even (2,2)-forms. We also turn on some $H_3$-flux on the basis of two odd 3-forms: 
\begin{equation}
    H_3 = p_ I b_ I \,.
\end{equation}

The superpotential that these fluxes generate is as follows: 
\begin{equation}
    W = 2 N_K p _ K + \frac{1}{4} e_0 + d_{ab} t _ a e_ b + \frac12 \kappa_{abc} t_a t_b m_ c + \frac{m_0}{6}\kappa _ { abc} t_a t _b t_c\,.
\end{equation}
From this, we can get the scalar potential from \eqref{eq:scalpot}. We start by writing the K\"ahler covariant derivatives of the superpotential:
\begin{align}\label{eq:Fterm}
    \text{Re}D_a W &= d_ {ab} e_ b + \kappa_{abc} u_b m_c + \frac{m_0}{2} \kappa_ {abc} ( u_b u_c - \nu_b \nu_ c) - \frac{\kappa_{abc}\nu_b\nu_c}{4 \mathcal{V}_6} \text{Im}W\,,\\
    \text{Im}D_a W &=  \kappa_{abc} \nu_b m_c + m_0 \kappa_ {abc} u_ b \nu_ c+ \frac{\kappa_{abc}\nu_b\nu_c}{4 \mathcal{V}_6} \text{Re}W\,,\\
    \text{Re}D_K W &= 2 p_ K - 4 i e^{ D} \mathcal{F}_K \text{Im}W\,,\\
    \text{Im}D_K W &= 4 i e^{ D} \mathcal{F}_K \text{Re}W\,. \label{eq:ImDKW_Fterm}
\end{align}
When equated to zero these give the F-term equations which leads to SUSY minima of the scalar potential. This was done explicitly in \cite{Ihl:2007ah}, and we summarize the relevant results below. \\

First, it is helpful to define the following quantities: 
\begin{equation}
    \hat{e}_1 = e_1 - \frac{m_2 m_3}{m_0}, \; \hat{e}_2= e_2 - \frac{m_1 m_3}{m_0},\; \hat{e}_3= e_3 - \frac{m_1 m_2 - 2 m_4^2}{m_0},\; \hat{e}_4=e_4 - \frac{m_3 m_4}{m_0}\,.
\end{equation}

For a sensible solution, the following conditions must be fullfilled:
\begin{align}
    &|p_1|>|p_2|\,, \\
    &\hat{e}_1 \hat{e}_2 > 2 \hat{e}_4^2\,,\\
   & m_0,\; \hat{e}_1,\;\hat{e}_2,\;\hat{e}_3 \text{ have the same sign}\,,\\
   & p_1 m_0 <0\,, \\
   & -\sqrt{2} m_0 p_1 + N_1 = 8\,,\\
   & - \sqrt{2} m_0 p_2 + N_2 = 0\,,
\end{align}
where the last two conditions come from the RR tadpole constraint and $N_i$ is the number of D6 branes wrapping the cycle dual to $b_i$.\\

The SUSY AdS vacua of the $\mathbb{T}^6/\mathbb{Z}_4$ model then have the following values for the scalar fields:
\begin{align}
    u_a &= -\frac{m_a}{m_0}\, , \quad U = \frac{1}{2}\frac{p_1+p_2}{p_1-p_2}\\
    \nu_1 &= |\hat{e}_2|\sqrt{\frac{5}{3m_0}\frac{\hat{e}_3}{\hat{e}_1\hat{e}_2-2\hat{e}_4^2}} \, , \quad \nu_2 = |\hat{e}_1|\sqrt{\frac{5}{3m_0}\frac{\hat{e}_3}{\hat{e}_1\hat{e}_2-2\hat{e}_4^2}}\\
    \nu_3 &= \sqrt{\frac{5}{3m_0}\frac{\hat{e}_1\hat{e}_2-2\hat{e}_4^2}{\hat{e}_3}} \, , \quad \nu_4 = \hat{e}_4\sqrt{\frac{5}{3m_0}\frac{\hat{e}_3}{\hat{e}_1\hat{e}_2-2\hat{e}_4^2}} \text{sign}(m_0)\\
    e^D &= \sqrt{\frac{27m_0}{10}\frac{p_1^2-p_2^2}{\hat{e}_3(\hat{e}_1\hat{e}_2-2\hat{e}_4^2)}}
\end{align}
Finally, one can do a field redefinition of the scalars $\xi_1$ and $\xi_2$,
\begin{align}
    \phi_1 &\equiv p_1 \xi_1 + p_2 \xi_2\, ,\\
    \phi_2 &\equiv p_2 \xi_1 - p_1 \xi_2\, .
\end{align}
The scalar potential is independent of $\phi_2$, so this direction remains flat perturbatively. Moreover, $\frac{\partial V}{\partial \phi_1}=0$ can be trivially solved for $\phi_1$ in terms of  the flux numbers and the fields $u_a$:
\begin{align}
    \phi_1 &= -\frac{1}{4} e_0 - d_{ab}u_a e_b - \frac{1}{2}\kappa_{abc}u_a u_b m_c - \frac{m_0}{6}\kappa_{abc} u_a u_b u_c \label{eq:phi1_in_terms_of_us}
\end{align}
Evaluated on the SUSY vacuum, this becomes 
\begin{align}
    \phi_1^{\text{vac}} &= -\frac{1}{4} e_0 + \frac{1}{m_0} d_{ab}m_a e_b - \frac{1}{3}\kappa_{abc}\frac{m_a m_b m_c}{m_0^2}
\end{align}
which is the same result as the one that could be obtained by solving the F-term equation  \eqref{eq:ImDKW_Fterm} for $\phi_1$ and plugging in the values of $u_a,\nu_a$ at the SUSY vacuum.

\subsection{The Black Hole Contribution}
Now we compute the gauge kinetic matrix which determines the black hole contribution. That one should be extra-ordinarily simple since there is only one vector. The general equation for the gauge kinetic matrix is
\begin{equation}
    f_{\alpha\beta}= i\hat{\kappa}_{\alpha\beta a}t^a\,,
\end{equation}
where
\begin{equation}
\hat{\kappa}_{\alpha\beta a} = \int_X \mu_{\alpha}\wedge \mu_{\beta}\wedge \omega_a\,.
\end{equation}
In our specific example there is just one vector, so no greek indices and we find that the only non-zero $\hat{\kappa}$ entry is in the third direction, $a=3$:
\begin{equation}
    \hat{\kappa}_3=-1\,.
\end{equation}
We rely on notation and conventions of \cite{Ihl:2007ah}. From this we find $f=-it_3=\nu_3 -iu_3$ leading  to the black hole potential (in the normalisation of appendix~\ref{app:Stability_BR_solutions})
\begin{align}
    V_{BH}(t_3) &= 
    \frac{1}{2\nu_3}\left(Q + P u_3\right)^2 + \frac{1}{2}P^2\nu_3\, .
\end{align}
We see that this potential on its own has no minimum at finite distance in moduli space: one can only obtain a singular (sometimes called \emph{small}) black hole (indeed $\partial_{u_3}V_{BH}=0$ gives $u_3=-Q/P$, after which $\partial_{\nu_3}V_{BH}=\frac12P^2\neq0$.). If one includes both the stabilizing potential and the BH potential, one can find a non-trivial minimum, as explained in the main text.

\section{Stability of Bertotti-Robinson solutions}
\label{app:Stability_BR_solutions}

Our starting point is the 4d action \eqref{eq:4d_SUGRA_action} (with only one gauge field, as in the main text):
\begin{equation}
    S= \int \dd^4 x \sqrt{-g}\left(\frac12 R   - K_{ I \bar J } \partial \phi ^I \partial\bar{\phi} ^{\bar J} - V - \frac14 (\text{Re} f(\phi)) F_{\mu\nu}  F^{\mu\nu} - \frac18 ( \text{Im} f(\phi)) \epsilon^{\mu\nu\rho\sigma}F_{\mu\nu} F_{\rho\sigma} \right)\,,
\end{equation}
We will proceed by expanding out the 4d EOMs around the metric ansatz 
\begin{align}
    \dd s_4^2 = \dd s_2^2 + e^{2\alpha \varphi(x)}\dd \Omega_2^2\, . \label{eq:4d_metric_ansatz_S2reduction}
\end{align}
and from there get the masses of the 12 real scalars and the volume modulus of the $S^2$, $\varphi(x)$. Note also the gauge field ansatz required to solve its EOM and Bianchi identity:
\begin{align}
    F &= P \dd \Omega_2 + \frac{e^{-2\alpha\varphi}}{\operatorname{Re}(f) }\left(\operatorname{Im}(f) P - Q\right) \vol_2 \label{eq:field_strength_4d_ansatz}
\end{align}

\subsection{Conventions}
Here we consider the subtleties associated to the change of conventions between the bottom-up and the top-down section of the paper: namely the $\frac{1}{2}\mathcal{R}$ or $\mathcal{R}$ term in the action. We show that the AdS$_2\times S^2$ equations are slightly different. 

The AdS$_2 \times S^2$ equations from our paper came from $S=\frac{1}{2 \kappa_4^2} \times$ 
\begin{align}
    \int \dd x^4 \sqrt{-g}\left(\mathcal{R}-\frac{1}{2} G_{i j}(\phi) \partial \phi^i \partial \phi^j-\frac{1}{4} A_{I J} F_{\mu \nu}^I F^{J \mu \nu}-\frac{1}{8} B_{I J} F_{\mu \nu}^I F_{\rho \sigma}^J \epsilon^{\mu \nu \rho \sigma}-V(\phi)\right)
\end{align}
and led to
\begin{align}
    &\partial_i V  + \frac{1}{R^4} \partial_i V_{BH} = 0\, ,\\
    &V - \frac{2}{R^2} + \frac{1}{R^4} V_{BH} = 0\, ,\\
    &-\frac{2}{R_{AdS}^2} = V-\frac{1}{R^4} V_{BH}\, . 
\end{align}
where
\begin{equation}
    V_{BH}(\phi) = \frac{1}{2}A P^2 + \frac{1}{2}A^{-1}\left(Q - B P\right)^2
\end{equation}
(where we adapted the notation to only 1 gauge field and used $P_I= B_{IJ}P^J$, $Q_I=Q$, $P^I=P$). Now our action is
\begin{align}
   \frac12 \int \dd^4 x \sqrt{-g}\left( R   - 2K_{ I \bar J } \partial \phi ^I \partial\bar{\phi} ^{\bar J} - 2V - 2\frac14 (\text{Re} f(\phi)) F_{\mu\nu}  F^{\mu\nu} - 2\frac18 ( \text{Im} f(\phi)) \epsilon^{\mu\nu\rho\sigma}F_{\mu\nu} F_{\rho\sigma} \right)\,,
\end{align}
So if we define $\phi=\frac{1}{2}\tilde{\phi}$,  identify (the real expansion of) $K$ with $G$ (i.e.\ $K_{I\bar J}\partial\phi^I\partial\bar\phi^{\bar J}\equiv G_{ij}\partial\phi^i\partial\phi^j$ in this subsection only; note that this $G$ is one half of the metric $G$ used in the rest of this appendix), set $\tilde{V}(\tilde{\phi})=2V(\phi)=2V(\tilde{\phi}/2)$ and rescale $F=\frac{1}{\sqrt{2}}\tilde{F}$, we get
\begin{align}
    S &= \int \dd^4 x \sqrt{-g}\left(\frac12 R   - K_{ I \bar J } \partial \phi ^I \partial\bar{\phi} ^{\bar J} - V - \frac14 (\text{Re} f(\phi)) F_{\mu\nu}  F^{\mu\nu} - \frac18 ( \text{Im} f(\phi)) \epsilon^{\mu\nu\rho\sigma}F_{\mu\nu} F_{\rho\sigma} \right)\,,\\
    &= \frac12 \int \dd^4 x \sqrt{-g}\left( R   - \frac{1}{2}G_{ij} \partial \tilde{\phi}^i \partial \tilde{\phi}^{j} - \tilde{V} - \frac14 A(\tilde{\phi}) \tilde{F}_{\mu\nu}  \tilde{F}^{\mu\nu} - \frac18 B(\tilde{\phi}) \epsilon^{\mu\nu\rho\sigma}\tilde{F}_{\mu\nu} \tilde{F}_{\rho\sigma} \right)\,,
\end{align}
where we also defined
\begin{align}
    \text{Re} f(\tilde{\phi}/2) = A(\tilde{\phi})\\
    \text{Im} f(\tilde{\phi}/2) = B(\tilde{\phi})
\end{align}
Note further that since $\tilde{F}=\sqrt{2}F$, the charges are related via $\tilde{Q}=\sqrt{2}Q$, and so $\tilde{V}_{BH}(\tilde{\phi})=2V_{BH}(\phi)=2V_{BH}(\tilde{\phi}/2)$ since it is quadratic in the charges. 

We now have the same conventions as in the first part of the paper and get the near-horizon equations in the tilde notation:
\begin{align}
    &\partial_{\tilde{\phi}^i} \tilde{V}  + \frac{1}{R^4} \partial_{\tilde{\phi}^i} \tilde{V}_{BH} = 0\, ,\\
    &\tilde{V} - \frac{2}{R^2} + \frac{1}{R^4} \tilde{V}_{BH} = 0\, ,\\
    &-\frac{2}{R_{AdS}^2} = \tilde{V}-\frac{1}{R^4} \tilde{V}_{BH}\, . 
\end{align}
Undoing the tilde notation leads to
\begin{align}
    &\partial_{i} V(\phi)  + \frac{1}{R^4} \partial_{i} V_{BH}(\phi)   = 0\, ,\\
    &V - \frac{1}{R^2} + \frac{1}{R^4} V_{BH} = 0\, ,\\
    &-\frac{1}{R_{AdS}^2} = V-\frac{1}{R^4} V_{BH}\, . \label{eq:lastequation}
\end{align}
Note the difference in factors of 2! The equations above can also be found directly from the Einstein equations, see below. The scalar equation is insensitive to the rescaling of $\phi$, and the last two equations imply $R_{AdS_2}^2=R^2/(1-2R^2V)$, i.e.\ \eqref{eq:RAdS2_eq_AdS2S2_top_down}.

\subsection{Equations of motion}

The 4d Einstein equations are
\begin{align}
    \mathcal{R}_{\mu\nu} = 2K_{ I \bar J } \partial_{\mu} \phi ^I \partial_{\nu} \bar{\phi} ^{\bar J} +  g_{\mu \nu}V -\frac{1}{4}  g_{\mu \nu}  (\text{Re} f(\phi)) F^2 + (\text{Re} f(\phi)) F_{\mu\rho} F_{\nu}^{\ \rho}
\end{align}
The Ricci tensor associated to the metric ansatz \eqref{eq:4d_metric_ansatz_S2reduction} splits as follows:
\begin{align}
    \mathcal{R}_{ab} &= \mathcal{R}^{(2)}_{ab} - 2\alpha\left(
    \nabla_a \partial_b \varphi
    +
    \alpha \partial_a \varphi\partial_b \varphi\right)\\
    \mathcal{R}_{ij} &= \bar g_{ij}\left(1-\alpha e^{2\alpha\varphi}\square\varphi - 2\alpha^2e^{2\alpha\varphi} (\partial\varphi)^2\right)
\end{align}
where $\bar{g}_{ij}$ is the metric on the 2-sphere without the warping. So the Einstein equation splits into the 2 parts as:
\begin{align}
    \mathcal{R}^{(2)}_{ab} - 2\alpha\left(
    \nabla_a \partial_b \varphi
    +
    \alpha \partial_a \varphi\partial_b \varphi\right) &= 2K_{ I \bar J } \partial_{a} \phi ^I \partial_{b} \bar{\phi} ^{\bar J} +  g_{ab} V -\frac{1}{4}  g_{ab}  (\text{Re} f(\phi)) F^2 + (\text{Re} f(\phi)) F_{a\rho} F_{b}^{\ \rho}\\
    \bar{g}_{ij}\left(1-\alpha e^{2\alpha\varphi}\square\varphi - 2\alpha^2e^{2\alpha\varphi} (\partial\varphi)^2\right) &= e^{2\alpha\varphi} \bar{g}_{ij}V -\frac{1}{4}  e^{2\alpha\varphi} \bar{g}_{ij} (\text{Re} f(\phi)) F^2 + (\text{Re} f(\phi)) F_{i\rho} F_{j}^{\ \rho}
\end{align}
Now, using our ansatz for the field strength \eqref{eq:field_strength_4d_ansatz}, the $S^2$ components of the Einstein equations become the EOM for the $S^2$ volume modulus:
\begin{align}
     \alpha \square_2\varphi + 2\alpha^2(\partial\varphi)^2 &= - V(\phi) + e^{-2\alpha\varphi} - e^{-4\alpha\varphi}V_{BH}(\phi) \, ,
\end{align}
where
\begin{align}
    V_{BH} &= \frac{1}{2}P^2 (\text{Re} f(\phi)) + \frac{1}{2\operatorname{Re}(f)}(P\operatorname{Im}(f)-Q)^2
\end{align}
This gives the BR equation \eqref{eq:R_eq_AdS2S2_top_down} when evaluating on constant scalars and using $R^2=e^{2\alpha\varphi}$. Moreover, when evaluating the external components of the Einstein equations on constant scalars, one gets 
\begin{align}
    \mathcal{R}^{(2)}_{ab} &= g_{ab}\left[ V
    - e^{-4\alpha\varphi}V_{BH}\right]
\end{align}
which gives the BR equation \eqref{eq:RAdS2_eq_AdS2S2_top_down}.\\

The scalar equations of motion are
\begin{align}
    \square_2 \phi^k
    +
    2\alpha \partial\varphi
    \partial \phi^k
    + \Gamma^{k}_{ij} \partial \phi^i \partial \phi^j 
    = G^{km}\left(\partial_m V
    + e^{-4\alpha\varphi}\partial_m V_{BH}\right)\, ,
\end{align}
where $G_{ij}$ is the real K\"ahler metric, normalised such that
\begin{align}
    K_{I\bar{J}} \partial \phi^I \partial \bar{\phi}^{\bar{J}} = \frac12 G_{ij}\partial \phi^i \partial \phi^j\,,
\end{align}
so that the scalar kinetic term is canonical and the geodesic distance in Planck units is measured by $\dd s^2=G_{ij}\dd\phi^i\dd\phi^j$. The $\Gamma^k_{ij}$ are the corresponding Christoffel symbols. Setting the scalars to be constants, we get back the BR equations \eqref{eq:NoBH_eq_AdS2S2_top_down} and \eqref{eq:t3_eq_AdS2S2_top_down}.\\

Before moving on to computing masses, note that from the form of the Kahler metric \eqref{eq:inverse_metric_moduli_space}, one can read off the form of $G^{ij}$: the matrix is block diagonal in the basis 
\begin{equation*}
\{u_1,u_2,u_4,u_3,\nu_1,\nu_2,\nu_4,\nu_3,U,D,\phi_1,\phi_2\}:
\end{equation*}
\begin{align}
    G^{ij} = \operatorname{Diag}(A,2\nu_3^2,A,2\nu_3^2,2U^2,\tfrac{1}{2},B) \label{eq:real_kinetic_matrix}
\end{align}
where $A$ is a 3x3 matrix and $B$ is a 2x2 matrix:
\begin{align}
    A &= \begin{pmatrix}
        2\nu_1^2 & 4\nu_4^2 & 2\nu_1\nu_4\\
        4\nu_4^2 & 2\nu_2^2 & 2\nu_2\nu_4\\
        2\nu_1\nu_4 & 2\nu_2\nu_4 & \frac{1}{2}\nu_1\nu_2 + \nu_4^2
    \end{pmatrix}\, , \\
    B &= \frac{e^{-2 D}}{4U}\begin{pmatrix}
    (p_1+p_2)^2 + 4 U^2 (p_1 - p_2)^2 & \left(p_1^2-p_2^2\right) \left(4 U^2-1\right) \\
     \left(p_1^2-p_2^2\right) \left(4 U^2-1\right)
     & (p_1-p_2)^2 + 4 U^2 (p_1 + p_2)^2 \\
    \end{pmatrix}
\end{align}

\subsection{Computing masses}
We now linearize the above scalar EOMs by setting $\phi^k=\phi_*^k+\delta^k$ and $\varphi=\varphi_* + \delta_{\varphi}$, where $\phi_*^k$ and $\varphi_*$ are the AdS$_2 \times S^2$ values of the scalars. One obtains:
\begin{align}
    \square_2 \delta_\varphi &= 
    2\frac{2 V_{BH}(\phi_*)-R_S^2}{R_S^4}\delta_{\varphi}\, , \\
    \square_2 \delta^k
    &=  G^{km}(\phi_*)\left(\partial_n \partial_m V + \frac{1}{R_S^4}\partial_n\partial_m V_{BH}\right) \delta^n
    -4\alpha \frac{1}{R_S^4}G^{km}(\phi_*)\partial_m V_{BH} \delta_{\varphi} \, .
\end{align}
where we wrote $e^{2\alpha\varphi_*}=R_S^2$. Note that the term proportional to $\delta^k$ in the linearized $\varphi$ equation vanishes when evaluated on the BR vacuum. Packaging our fields as $\vec{\delta}=(\delta_{\varphi},\delta^k)$ we get
\begin{align}
    \square_2 \vec{\delta} = \begin{pmatrix}
        2\frac{2 V_{BH}(\phi_*)-R_S^2}{R_S^4} & 0\\
        -4\alpha \frac{1}{R_S^4}G^{km}(\phi_*)\partial_m V_{BH}  & G^{km}(\phi_*)\left(\partial_m \partial_n V + \frac{1}{R_S^4}\partial_m\partial_n V_{BH}\right)
    \end{pmatrix}\vec{\delta} \label{eq:AdS2_mass_matrix}
\end{align}
and the masses (squared) are simply the eigenvalues of this matrix.

\section{Scalar Field Distances in Fuzzballs, by Daniel Mayerson} \label{App:FUZZ} 
In this section, we discuss how near the horizon of a fuzzball, towers of states become light at the horizon, as the scalar fields explore the boundary of moduli space (see \cite{Li:2021utg} for related work). Fuzzballs are stringy geometries which asymptotically look exactly the same as a black hole. They play an important role in the fuzzball proposal for the resolution of the information problem (see e.g. \cite{Mathur:2005zp} for a review). The intuition behind fuzzballs is that they resolve the black hole interior into smooth higher-dimensional geometries. From the perspective of the EFT this gives some intuition as to why they should probe infinite distance limits in moduli space: near the horizon and ``inside'' the black hole, the ten (or eleven) dimensional picture becomes the relevant one. We now argue for this in a four dimensional example. 

We work with the 4D action:
\be S = \frac{1}{16\pi G} \int\left( * R - 2\sum_I \frac{dz^I\wedge d\bar{z}^I}{(\Im(z^I))^2} + \cdots\right),\ee
where $I=1,2,3$, and the ellipses represent the rest of the field content of the theory. One can simplify further and set everything with an $I$ index equal, so $z^I = z$. We are interested in the multi-centered solutions which are supersymmetric and have a metric given by:
\be ds^2 = - \Sigma^{-1} (dt+\omega)^2 + \Sigma ds_3^2,\ee
These are called Denef or Denef-Bates geometries, and we use the Bena-Warner 5D conventions. The 3D metric $ds_3^2$ is just the flat metric on $\mathbb{R}^3$ and:
\be \Sigma^2 = Z^3V - \mu^2 V^2, \quad Z = L + \frac{K^2}{V}, \quad \mu = M +\frac{3K L}{2V} + \frac{K^3}{V^2},\ee
where $V,K,L,M$ are harmonic functions that determine the entire solution. 
The rotation form $\omega$ can also be determined explicitly but we will not need it here. The scalars are given by:
 \be z^ I = z = \frac{K + 2 i \frac{\partial \Sigma}{\partial L}}{V + i \frac{\partial \Sigma}{\partial M}}.\ee

\paragraph{Black hole}
Let us first consider the black hole geometry given by:
\be V = L = 0, \qquad K = 1 + 2\frac{P}{r}, \qquad M = -\frac12 + \frac{(P^3 - q_0)}{r}.\ee
We will always take $(r,\theta,\phi)$ or $(x,y,z)$ to be the standard coordinates on the (flat) $\mathbb{R}^3$ basis. Note that in order to get a sensible answer for $\Sigma$, first take $V=\epsilon$ and then take $\epsilon\rightarrow 0$ in the expression above. This is a black hole with horizon at $r=0$, with a finite area:
\be A_{\text{BH}} = 4 \sqrt{P^3 \left(q_0-P^3\right)},\ee
so we see that $P>0$ and $q_0>P^3$ is sufficient to keep this black hole well behaved. The scalar goes to a finite value at the horizon and moreover vanishes quickly enough at infinity, to make the total moduli space distance traveled by the scalar nicely finite:
\be \int_{0}^{\infty} dr \sqrt{\frac{\partial_r z \partial_r \bar{z}}{(\Im(z))^2}} = \frac12 \log\left( \frac{P}{q_0-P^3}\right).\ee

\paragraph{Smooth centers} We now describe a what is known as a smooth center, which will be the fundamental building block for the fuzzball. A smooth center is a  fluxed D6 brane wrapping the internal space and is given by taking:
\be V \sim \frac{1}{r},\quad K\sim \frac{P}{r}, \quad L\sim \frac{-P^2}{r}, \quad M\sim \frac{P^3/2}{r},\ee
where $\sim$ denotes that only the singularity at $r=0$ (where the D6 brane is) is shown. Although this is singular from the perspective of (super-)gravity, it is smooth in the sense that in string theory it is resolved by the open strings that end on the D6.

It is easy to see that a smooth center will always give divergences in the scalar distance traveled. It is sufficient to know that the scalar blows up as:
\be z \sim c_1 + i c_2 \sqrt{r},\ee
when approaching such a smooth center at $r\rightarrow 0$. It follows that $\partial_r z \sim 1/\sqrt{r}$ and $\Im(z)\sim \sqrt{r}$ so that:
\be \sqrt{\frac{|\partial_r z|^2}{(\Im(z))^2}} \sim \sqrt{ \frac{1}{r} \frac{1}{r} } \sim \frac{1}{r},\ee
whose integral diverges.

\paragraph{Scaling solution} We now describe a fuzzball that corresponds to the black hole solution above. The analysis above certainly holds for one or two smooth centers. If we have more than two, we can build so-called scaling solutions such as:
\be V = \frac{1}{r_1} - \frac{1}{r_2}, \quad K = 1 + P\left(\frac{1}{r_1}+\frac{1}{r_2}\right), \quad L = -P^2\left(\frac{1}{r_1} - \frac{1}{r_2}\right), \quad M = -\frac12 + \frac{P^3}{2}\left(\frac{1}{r_1} + \frac{1}{r_2}\right) -\frac{q_0}{r_3}.\ee
This represents a D6 and anti-D6 with fluxes, plus some number $q_0$ of D0 branes (at a different location). The $r_i$ are the distances to the particular center; we will take the centers to lie at $x=y=0$ and $z=\pm l/2$ (for the D6 and anti-D6) and the third (D0) center at $y=z=0$ and $x=R$. These parameters $l,R$ are determined by the bubble equations (typically called integrability equations in 4D):
\be l = 8P^3 \lambda, \quad R = 2\lambda \sqrt{\frac{q_0^2}{(1-(1-3P^2)\lambda)^2 - 4P^6}}.\ee
The parameter $\lambda$ is a (semi-)free parameter and is called the scaling parameter.
It is certainly instructive to figure out what values of $q_0,P,\lambda$ are allowed, and this was done in  \cite{Bacchini:2021fig}. One can for instance pick $P=2$ and $q_0=50$ which are allowed values and give an allowed range for $\lambda$ of $0 < \lambda < 17/88\approx 0.193$.

As we decrease the scaling parameter $\lambda$, the coordinate distance between the centers becomes smaller. Physically, the proper distance between the centers asymptotes to a fixed, finite value, and rather what happens is that a deeper and deeper extremal-black-hole-like $AdS_2$ throat emerges as $\lambda$ decreases. The strict $\lambda\rightarrow 0$ limit corresponds to precisely the BH we discussed earlier (with a finite-sized horizon). This is the essence of a fuzzball geometry: it can be made to resemble the corresponding BH arbitrarily well as we approach the so-called scaling point $\lambda\rightarrow 0$.

We now turn to the scalars: what happens to the moduli space distance traveled by the scalars as we play with $\lambda$? A series expansion of the scalar near one of the D6 branes reveals:
\be z = P + i \sqrt{r_1} \sqrt{\frac{1+4 \lambda  P^2}{2 P \lambda}} + \mathcal{O}(r_1),\ee
so the moduli space distance traveled by the scalars always diverges, and moreover the behaviour to leading order in $\lambda$ is:
\be z \sim i \sqrt{\frac{r_1}{\lambda}},\ee
so that:
\be \sqrt{\frac{|\partial_r z|^2}{(\Im(z))^2}} \sim \sqrt{ \frac{1}{\lambda\, r_1} \frac{\lambda}{r_1} } \sim \frac{1}{r_1}.\ee
This shows that the divergence of the scalar moduli space distance is (to leading order in $\lambda$) actually \emph{independent} of the scaling parameter $\lambda$ --- meaning that the divergence carries on all the way to $\lambda\rightarrow 0$. Clearly, this is another interesting way in which the $\lambda\rightarrow 0$ limit is discontinuous, as the corresponding BH (``at'' $\lambda=0$) has a finite moduli space distance traveled by the scalars, as calculated above.

\bibliographystyle{utphys}
\bibliography{refs}

@article{Anabalon:2013qua,
    author = "Anabalon, Andres and Astefanesei, Dumitru and Mann, Robert",
    title = "{Exact asymptotically flat charged hairy black holes with a dilaton potential}",
    eprint = "1308.1693",
    archivePrefix = "arXiv",
    primaryClass = "hep-th",
    doi = "10.1007/JHEP10(2013)184",
    journal = "JHEP",
    volume = "10",
    pages = "184",
    year = "2013"
}

@article{Benakli:2026zbm,
    author = "Benakli, Karim and Chrysostomou, Anna",
    title = "{The Fate of Black Hole-Induced Moduli Excursions in the Presence of Scalar Potentials}",
    eprint = "2607.05488",
    archivePrefix = "arXiv",
    primaryClass = "hep-th",
    month = "7",
    year = "2026"
}

@article{Gregory:1992kr,
    author = "Gregory, Ruth and Harvey, Jeffrey A.",
    title = "{Black holes with a massive dilaton}",
    eprint = "hep-th/9209070",
    archivePrefix = "arXiv",
    reportNumber = "EFI-92-49",
    doi = "10.1103/PhysRevD.47.2411",
    journal = "Phys. Rev. D",
    volume = "47",
    pages = "2411--2422",
    year = "1993"
}

@article{Skenderis:2008qn,
    author = "Skenderis, Kostas and Taylor, Marika",
    title = "{The fuzzball proposal for black holes}",
    eprint = "0804.0552",
    archivePrefix = "arXiv",
    primaryClass = "hep-th",
    reportNumber = "ITFA-2008-12",
    doi = "10.1016/j.physrep.2008.08.001",
    journal = "Phys. Rept.",
    volume = "467",
    pages = "117--171",
    year = "2008"
}

@article{DallAgata:2011zkh,
    author = "Dall'Agata, Gianguido",
    editor = "Bellucci, Stefano",
    title = "{Black holes in supergravity: flow equations and duality}",
    eprint = "1106.2611",
    archivePrefix = "arXiv",
    primaryClass = "hep-th",
    reportNumber = "DFPD-11-TH-07",
    doi = "10.1007/978-3-642-31380-6_1",
    journal = "Springer Proc. Phys.",
    volume = "142",
    pages = "1--45",
    year = "2013"
}

@article{Dabholkar:2012zz,
    author = "Dabholkar, Atish and Nampuri, Suresh",
    title = "{Quantum black holes}",
    eprint = "1208.4814",
    archivePrefix = "arXiv",
    primaryClass = "hep-th",
    doi = "10.1007/978-3-642-25947-0_5",
    journal = "Lect. Notes Phys.",
    volume = "851",
    pages = "165--232",
    year = "2012"
}

@article{Lee:2019wij,
    author = "Lee, Seung-Joo and Lerche, Wolfgang and Weigand, Timo",
    title = "{Emergent strings from infinite distance limits}",
    eprint = "1910.01135",
    archivePrefix = "arXiv",
    primaryClass = "hep-th",
    reportNumber = "CERN-TH-2019-159",
    doi = "10.1007/JHEP02(2022)190",
    journal = "JHEP",
    volume = "02",
    pages = "190",
    year = "2022"
}

@article{Mathur:2005zp,
    author = "Mathur, Samir D.",
    editor = "Kiritsis, E.",
    title = "{The Fuzzball proposal for black holes: An Elementary review}",
    eprint = "hep-th/0502050",
    archivePrefix = "arXiv",
    doi = "10.1002/prop.200410203",
    journal = "Fortsch. Phys.",
    volume = "53",
    pages = "793--827",
    year = "2005"
}

@article{Lust:2022lfc,
    author = {L\"ust, Severin and Vafa, Cumrun and Wiesner, Max and Xu, Kai},
    title = "{Holography and the KKLT scenario}",
    eprint = "2204.07171",
    archivePrefix = "arXiv",
    primaryClass = "hep-th",
    doi = "10.1007/JHEP10(2022)188",
    journal = "JHEP",
    volume = "10",
    pages = "188",
    year = "2022"
}

@article{Demirtas:2021nlu,
    author = "Demirtas, Mehmet and Kim, Manki and McAllister, Liam and Moritz, Jakob and Rios-Tascon, Andres",
    title = "{Small cosmological constants in string theory}",
    eprint = "2107.09064",
    archivePrefix = "arXiv",
    primaryClass = "hep-th",
    doi = "10.1007/JHEP12(2021)136",
    journal = "JHEP",
    volume = "12",
    pages = "136",
    year = "2021"
}

@article{Giddings:2001yu,
    author = "Giddings, Steven B. and Kachru, Shamit and Polchinski, Joseph",
    title = "{Hierarchies from fluxes in string compactifications}",
    eprint = "hep-th/0105097",
    archivePrefix = "arXiv",
    reportNumber = "SLAC-PUB-8807, NSF-ITP-01-37, SU-ITP-01-16",
    doi = "10.1103/PhysRevD.66.106006",
    journal = "Phys. Rev. D",
    volume = "66",
    pages = "106006",
    year = "2002"
}

@article{Bobev:2023dwx,
    author = "Bobev, Nikolay and David, Marina and Hong, Junho and Reys, Valentin and Zhang, Xuao",
    title = "{A compendium of logarithmic corrections in AdS/CFT}",
    eprint = "2312.08909",
    archivePrefix = "arXiv",
    primaryClass = "hep-th",
    doi = "10.1007/JHEP04(2024)020",
    journal = "JHEP",
    volume = "04",
    pages = "020",
    year = "2024"
}

@article{Kachru:2003aw,
    author = "Kachru, Shamit and Kallosh, Renata and Linde, Andrei D. and Trivedi, Sandip P.",
    title = "{De Sitter vacua in string theory}",
    eprint = "hep-th/0301240",
    archivePrefix = "arXiv",
    reportNumber = "SLAC-PUB-9630, SU-ITP-03-01, TIFR-TH-03-03",
    doi = "10.1103/PhysRevD.68.046005",
    journal = "Phys. Rev. D",
    volume = "68",
    pages = "046005",
    year = "2003"
}

@article{Balasubramanian:2005zx,
    author = "Balasubramanian, Vijay and Berglund, Per and Conlon, Joseph P. and Quevedo, Fernando",
    title = "{Systematics of moduli stabilisation in Calabi-Yau flux compactifications}",
    eprint = "hep-th/0502058",
    archivePrefix = "arXiv",
    reportNumber = "DAMTP-2005-10, UNH-05-01, UPR-1109-T",
    doi = "10.1088/1126-6708/2005/03/007",
    journal = "JHEP",
    volume = "03",
    pages = "007",
    year = "2005"
}

@article{Ferrara:1996dd,
    author = "Ferrara, Sergio and Kallosh, Renata",
    title = "{Supersymmetry and attractors}",
    eprint = "hep-th/9602136",
    archivePrefix = "arXiv",
    reportNumber = "UCLA-96-TEP-8, CERN-TH-96-53, SU-ITP-96-8",
    doi = "10.1103/PhysRevD.54.1514",
    journal = "Phys. Rev. D",
    volume = "54",
    pages = "1514--1524",
    year = "1996"
}

@article{Denef:2000nb,
    author = "Denef, Frederik",
    title = "{Supergravity flows and D-brane stability}",
    eprint = "hep-th/0005049",
    archivePrefix = "arXiv",
    doi = "10.1088/1126-6708/2000/08/050",
    journal = "JHEP",
    volume = "08",
    pages = "050",
    year = "2000"
}

@article{Cheung:2018cwt,
    author = "Cheung, Clifford and Liu, Junyu and Remmen, Grant N.",
    title = "{Proof of the Weak Gravity Conjecture from Black Hole Entropy}",
    eprint = "1801.08546",
    archivePrefix = "arXiv",
    primaryClass = "hep-th",
    reportNumber = "CALT-TH-2018-007",
    doi = "10.1007/JHEP10(2018)004",
    journal = "JHEP",
    volume = "10",
    pages = "004",
    year = "2018"
}

@article{Arkani-Hamed:2006emk,
    author = "Arkani-Hamed, Nima and Motl, Lubos and Nicolis, Alberto and Vafa, Cumrun",
    title = "{The String landscape, black holes and gravity as the weakest force}",
    eprint = "hep-th/0601001",
    archivePrefix = "arXiv",
    reportNumber = "HUTP-05-A0057",
    doi = "10.1088/1126-6708/2007/06/060",
    journal = "JHEP",
    volume = "06",
    pages = "060",
    year = "2007"
}

@article{Grana:2006kf,
    author = "Grana, Mariana and Minasian, Ruben and Petrini, Michela and Tomasiello, Alessandro",
    title = "{A Scan for new N=1 vacua on twisted tori}",
    eprint = "hep-th/0609124",
    archivePrefix = "arXiv",
    reportNumber = "SPHT-T06-094, SU-ITP-06-23",
    doi = "10.1088/1126-6708/2007/05/031",
    journal = "JHEP",
    volume = "05",
    pages = "031",
    year = "2007"
}

@article{Almheiri:2020cfm,
    author = "Almheiri, Ahmed and Hartman, Thomas and Maldacena, Juan and Shaghoulian, Edgar and Tajdini, Amirhossein",
    title = "{The entropy of Hawking radiation}",
    eprint = "2006.06872",
    archivePrefix = "arXiv",
    primaryClass = "hep-th",
    doi = "10.1103/RevModPhys.93.035002",
    journal = "Rev. Mod. Phys.",
    volume = "93",
    number = "3",
    pages = "035002",
    year = "2021"
}

@article{Strominger:1996sh,
    author = "Strominger, Andrew and Vafa, Cumrun",
    title = "{Microscopic origin of the Bekenstein-Hawking entropy}",
    eprint = "hep-th/9601029",
    archivePrefix = "arXiv",
    reportNumber = "HUTP-96-A002, RU-96-01",
    doi = "10.1016/0370-2693(96)00345-0",
    journal = "Phys. Lett. B",
    volume = "379",
    pages = "99--104",
    year = "1996"
}

@article{Vafa:2005ui,
    author = "Vafa, Cumrun",
    title = "{The String landscape and the swampland}",
    eprint = "hep-th/0509212",
    archivePrefix = "arXiv",
    reportNumber = "HUTP-05-A043",
    month = "9",
    year = "2005"
}

@article{Coudarchet:2023mfs,
    author = "Coudarchet, Thibaut",
    title = "{Hiding the extra dimensions: A review on scale separation in string theory}",
    eprint = "2311.12105",
    archivePrefix = "arXiv",
    primaryClass = "hep-th",
    doi = "10.1016/j.physrep.2024.02.003",
    journal = "Phys. Rept.",
    volume = "1064",
    pages = "1--28",
    year = "2024"
}

@article{Sen:2012dw,
    author = "Sen, Ashoke",
    title = "{Logarithmic Corrections to Schwarzschild and Other Non-extremal Black Hole Entropy in Different Dimensions}",
    eprint = "1205.0971",
    archivePrefix = "arXiv",
    primaryClass = "hep-th",
    doi = "10.1007/JHEP04(2013)156",
    journal = "JHEP",
    volume = "04",
    pages = "156",
    year = "2013"
}

@article{Mathur:2009hf,
    author = "Mathur, Samir D.",
    editor = "Uranga, A. M.",
    title = "{The Information paradox: A Pedagogical introduction}",
    eprint = "0909.1038",
    archivePrefix = "arXiv",
    primaryClass = "hep-th",
    doi = "10.1088/0264-9381/26/22/224001",
    journal = "Class. Quant. Grav.",
    volume = "26",
    pages = "224001",
    year = "2009"
}

@article{Acharya:2006ne,
    author = "Acharya, Bobby Samir and Benini, Francesco and Valandro, Roberto",
    title = "{Fixing moduli in exact type IIA flux vacua}",
    eprint = "hep-th/0607223",
    archivePrefix = "arXiv",
    doi = "10.1088/1126-6708/2007/02/018",
    journal = "JHEP",
    volume = "02",
    pages = "018",
    year = "2007"
}

@article{Cribiori:2021djm,
    author = "Cribiori, Niccol\`o and Junghans, Daniel and Van Hemelryck, Vincent and Van Riet, Thomas and Wrase, Timm",
    title = "{Scale-separated AdS4 vacua of IIA orientifolds and M-theory}",
    eprint = "2107.00019",
    archivePrefix = "arXiv",
    primaryClass = "hep-th",
    reportNumber = "UUITP-29/21",
    doi = "10.1103/PhysRevD.104.126014",
    journal = "Phys. Rev. D",
    volume = "104",
    number = "12",
    pages = "126014",
    year = "2021"
}

@article{Junghans:2020acz,
    author = "Junghans, Daniel",
    title = "{O-Plane Backreaction and Scale Separation in Type IIA Flux Vacua}",
    eprint = "2003.06274",
    archivePrefix = "arXiv",
    primaryClass = "hep-th",
    doi = "10.1002/prop.202000040",
    journal = "Fortsch. Phys.",
    volume = "68",
    number = "6",
    pages = "2000040",
    year = "2020"
}

@article{Montero:2024qtz,
    author = "Montero, Miguel and Valenzuela, Irene",
    title = "{Quantum corrections to DGKT and the Weak Gravity Conjecture}",
    eprint = "2412.00189",
    archivePrefix = "arXiv",
    primaryClass = "hep-th",
    reportNumber = "IFT-24-155, CERN-TH-2024-190",
    month = "11",
    year = "2024"
}

@article{Elias-Miro:2011sqh,
    author = "Elias-Miro, Joan and Espinosa, Jose R. and Giudice, Gian F. and Isidori, Gino and Riotto, Antonio and Strumia, Alessandro",
    title = "{Higgs mass implications on the stability of the electroweak vacuum}",
    eprint = "1112.3022",
    archivePrefix = "arXiv",
    primaryClass = "hep-ph",
    doi = "10.1016/j.physletb.2012.02.013",
    journal = "Phys. Lett. B",
    volume = "709",
    pages = "222--228",
    year = "2012"
}

@article{Danielsson:2017riq,
    author = "Danielsson, U. H. and Dibitetto, G. and Giri, S.",
    title = "{Black holes as bubbles of AdS}",
    eprint = "1705.10172",
    archivePrefix = "arXiv",
    primaryClass = "hep-th",
    reportNumber = "UUITP-15-17",
    doi = "10.1007/JHEP10(2017)171",
    journal = "JHEP",
    volume = "10",
    pages = "171",
    year = "2017"
}

@article{Palti:2019pca,
    author = "Palti, Eran",
    title = "{The Swampland: Introduction and Review}",
    eprint = "1903.06239",
    archivePrefix = "arXiv",
    primaryClass = "hep-th",
    reportNumber = "MPP-2019-53",
    doi = "10.1002/prop.201900037",
    journal = "Fortsch. Phys.",
    volume = "67",
    number = "6",
    pages = "1900037",
    year = "2019"
}

@article{Basile:2023blg,
    author = {Basile, Ivano and L\"ust, Dieter and Montella, Carmine},
    title = "{Shedding black hole light on the emergent string conjecture}",
    eprint = "2311.12113",
    archivePrefix = "arXiv",
    primaryClass = "hep-th",
    reportNumber = "LMU-ASC 35/23, MPP-2023-262",
    doi = "10.1007/JHEP07(2024)208",
    journal = "JHEP",
    volume = "07",
    pages = "208",
    year = "2024"
}

@article{Sen:2025bmj,
    author = "Sen, Ashoke",
    title = "{Are Moduli Vacuum Expectation Values or Parameters?}",
    eprint = "2502.07883",
    archivePrefix = "arXiv",
    primaryClass = "hep-th",
    month = "2",
    year = "2025"
}

@article{Banks:2025nfe,
    author = "Banks, T.",
    title = "{Old Ideas for New Physicists III: String Theory Parameters are NOT Vacuum Expectation Values}",
    eprint = "2501.17697",
    archivePrefix = "arXiv",
    primaryClass = "hep-th",
    reportNumber = "RUNHETC-2025-7",
    month = "1",
    year = "2025"
}

@article{Emelin:2024vug,
    author = "Emelin, Maxim",
    title = "{Consistency conditions for O-plane unsmearing from second-order perturbation theory}",
    eprint = "2407.12717",
    archivePrefix = "arXiv",
    primaryClass = "hep-th",
    doi = "10.1007/JHEP12(2024)025",
    journal = "JHEP",
    volume = "12",
    pages = "025",
    year = "2024"
}

@article{Calderon-Infante:2023uhz,
    author = "Calder\'on-Infante, Jos\'e and Delgado, Matilda and Uranga, Angel M.",
    title = "{Emergence of species scale black hole horizons}",
    eprint = "2310.04488",
    archivePrefix = "arXiv",
    primaryClass = "hep-th",
    reportNumber = "IFT-UAM/CSIC-23-123, CERN-TH-2023-173",
    doi = "10.1007/JHEP01(2024)003",
    journal = "JHEP",
    volume = "01",
    pages = "003",
    year = "2024"
}

@article{Ooguri:2018wrx,
    author = "Ooguri, Hirosi and Palti, Eran and Shiu, Gary and Vafa, Cumrun",
    title = "{Distance and de Sitter Conjectures on the Swampland}",
    eprint = "1810.05506",
    archivePrefix = "arXiv",
    primaryClass = "hep-th",
    doi = "10.1016/j.physletb.2018.11.018",
    journal = "Phys. Lett. B",
    volume = "788",
    pages = "180--184",
    year = "2019"
}

@article{VanRiet:2023pnx,
    author = "Van Riet, Thomas and Zoccarato, Gianluca",
    title = "{Beginners lectures on flux compactifications and related Swampland topics}",
    eprint = "2305.01722",
    archivePrefix = "arXiv",
    primaryClass = "hep-th",
    doi = "10.1016/j.physrep.2023.11.003",
    journal = "Phys. Rept.",
    volume = "1049",
    pages = "1--51",
    year = "2024"
}

@article{Baines:2020dmu,
    author = "Baines, Stephanie and Van Riet, Thomas",
    title = "{Smearing orientifolds in flux compactifications can be OK}",
    eprint = "2005.09501",
    archivePrefix = "arXiv",
    primaryClass = "hep-th",
    doi = "10.1088/1361-6382/aba8e0",
    journal = "Class. Quant. Grav.",
    volume = "37",
    number = "19",
    pages = "195015",
    year = "2020"
}

@article{Marchesano:2020qvg,
    author = "Marchesano, Fernando and Palti, Eran and Quirant, Joan and Tomasiello, Alessandro",
    title = "{On supersymmetric AdS$_{4}$ orientifold vacua}",
    eprint = "2003.13578",
    archivePrefix = "arXiv",
    primaryClass = "hep-th",
    reportNumber = "IFT-UAM/CSIC-20-51",
    doi = "10.1007/JHEP08(2020)087",
    journal = "JHEP",
    volume = "08",
    pages = "087",
    year = "2020"
}

@article{Rajaguru:2024emw,
    author = "Rajaguru, Muthusamy and Sengupta, Anindya and Wrase, Timm",
    title = "{Fully stabilized Minkowski vacua in the 2$^{6}$ Landau-Ginzburg model}",
    eprint = "2407.16756",
    archivePrefix = "arXiv",
    primaryClass = "hep-th",
    doi = "10.1007/JHEP10(2024)095",
    journal = "JHEP",
    volume = "10",
    pages = "095",
    year = "2024"
}

@article{Lust:2019zwm,
    author = {L\"ust, Dieter and Palti, Eran and Vafa, Cumrun},
    title = "{AdS and the Swampland}",
    eprint = "1906.05225",
    archivePrefix = "arXiv",
    primaryClass = "hep-th",
    doi = "10.1016/j.physletb.2019.134867",
    journal = "Phys. Lett. B",
    volume = "797",
    pages = "134867",
    year = "2019"
}

@article{Gautason:2018gln,
    author = "Gautason, F. F. and Van Hemelryck, V. and Van Riet, T.",
    title = "{The Tension between 10D Supergravity and dS Uplifts}",
    eprint = "1810.08518",
    archivePrefix = "arXiv",
    primaryClass = "hep-th",
    doi = "10.1002/prop.201800091",
    journal = "Fortsch. Phys.",
    volume = "67",
    number = "1-2",
    pages = "1800091",
    year = "2019"
}

@article{Calderon-Infante:2025ldq,
    author = "Calder\'on-Infante, Jos\'e and Castellano, Alberto and Herr\'aez, Alvaro",
    title = "{The Double EFT Expansion in Quantum Gravity}",
    eprint = "2501.14880",
    archivePrefix = "arXiv",
    primaryClass = "hep-th",
    month = "1",
    year = "2025"
}

@article{Castellano:2025ljk,
    author = "Castellano, Alberto and Zatti, Matteo",
    title = "{Black Hole Entropy, Quantum Corrections and EFT Transitions}",
    eprint = "2502.02655",
    archivePrefix = "arXiv",
    primaryClass = "hep-th",
    month = "2",
    year = "2025"
}

@article{Dvali:2007hz,
      author         = "Dvali, Gia",
      title          = "{Black Holes and Large N Species Solution to the
                        Hierarchy Problem}",
      journal        = "Fortsch. Phys.",
      volume         = "58",
      year           = "2010",
      pages          = "528-536",
      doi            = "10.1002/prop.201000009",
      eprint         = "0706.2050",
      archivePrefix  = "arXiv",
      primaryClass   = "hep-th",
      SLACcitation   = "%%CITATION = ARXIV:0706.2050;%%"
}

@article{Dvali:2007wp,
    author = "Dvali, Gia and Redi, Michele",
    title = "{Black Hole Bound on the Number of Species and Quantum Gravity at LHC}",
    eprint = "0710.4344",
    archivePrefix = "arXiv",
    primaryClass = "hep-th",
    doi = "10.1103/PhysRevD.77.045027",
    journal = "Phys. Rev. D",
    volume = "77",
    pages = "045027",
    year = "2008"
}

@article{Dvali:2008ec,
    author = "Dvali, Gia and Gomez, Cesar",
    title = "{Quantum Information and Gravity Cutoff in Theories with Species}",
    eprint = "0812.1940",
    archivePrefix = "arXiv",
    primaryClass = "hep-th",
    reportNumber = "CERN-PH-TH-2008-226, IFT-UAM-CSIC-08-79",
    doi = "10.1016/j.physletb.2009.03.024",
    journal = "Phys. Lett. B",
    volume = "674",
    pages = "303--307",
    year = "2009"
}

@article{Dvali:2009ks,
    author = "Dvali, Gia and L{\"u}st, Dieter",
    title = "{Evaporation of Microscopic Black Holes in String Theory and the Bound on Species}",
    eprint = "0912.3167",
    archivePrefix = "arXiv",
    primaryClass = "hep-th",
    reportNumber = "CERN-PH-TH-2009-243, MPP-2009-205, LMU-ASC-56-09",
    doi = "10.1002/prop.201000008",
    journal = "Fortsch. Phys.",
    volume = "58",
    pages = "505--527",
    year = "2010"
}

@article{Dvali:2010vm,
    author = "Dvali, Gia and Gomez, Cesar",
    title = "{Species and Strings}",
    eprint = "1004.3744",
    archivePrefix = "arXiv",
    primaryClass = "hep-th",
    reportNumber = "CERN-PH-TH-2010-069, IFT-UAM-CSIC-10-25",
    month = "4",
    year = "2010"
}

@article{vandeHeisteeg:2023dlw,
    author = "van de Heisteeg, Damian and Vafa, Cumrun and Wiesner, Max and Wu, David H.",
    title = "{Species Scale in Diverse Dimensions}",
    eprint = "2310.07213",
    archivePrefix = "arXiv",
    primaryClass = "hep-th",
    month = "10",
    year = "2023"
}

@article{Castellano:2023aum,
    author = "Castellano, Alberto and Herr\'aez, Alvaro and Ib\'a\~nez, Luis E.",
    title = "{On the species scale, modular invariance and the gravitational EFT expansion}",
    eprint = "2310.07708",
    archivePrefix = "arXiv",
    primaryClass = "hep-th",
    doi = "10.1007/JHEP12(2024)019",
    journal = "JHEP",
    volume = "12",
    pages = "019",
    year = "2024"
}

@article{Bedroya:2024ubj,
    author = "Bedroya, Alek and Mishra, Rashmish K. and Wiesner, Max",
    title = "{Density of States, Black Holes and the Emergent String Conjecture}",
    eprint = "2405.00083",
    archivePrefix = "arXiv",
    primaryClass = "hep-th",
    month = "4",
    year = "2024"
}

@article{Castellano:2022bvr,
    author = "Castellano, Alberto and Herr\'aez, Alvaro and Ib\'a\~nez, Luis E.",
    title = "{The emergence proposal in quantum gravity and the species scale}",
    eprint = "2212.03908",
    archivePrefix = "arXiv",
    primaryClass = "hep-th",
    reportNumber = "IFT-UAM/CSIC-22-149",
    doi = "10.1007/JHEP06(2023)047",
    journal = "JHEP",
    volume = "06",
    pages = "047",
    year = "2023"
}

@article{vandeHeisteeg:2022btw,
    author = "van de Heisteeg, Damian and Vafa, Cumrun and Wiesner, Max and Wu, David H.",
    title = "{Moduli-dependent Species Scale}",
    eprint = "2212.06841",
    archivePrefix = "arXiv",
    primaryClass = "hep-th",
    month = "12",
    year = "2022"
}

@article{Cribiori:2022nke,
    author = {Cribiori, Niccol\`o and L{\"u}st, Dieter and Staudt, Georgina},
    title = "{Black hole entropy and moduli-dependent species scale}",
    eprint = "2212.10286",
    archivePrefix = "arXiv",
    primaryClass = "hep-th",
    reportNumber = "LMU-ASC 56/22, MPP-2022-289",
    doi = "10.1016/j.physletb.2023.138113",
    journal = "Phys. Lett. B",
    volume = "844",
    pages = "138113",
    year = "2023"
}

@article{Dvali:2012uq,
    author = "Dvali, Gia and Gomez, Cesar and L{\"u}st, Dieter",
    title = "{Black Hole Quantum Mechanics in the Presence of Species}",
    eprint = "1206.2365",
    archivePrefix = "arXiv",
    primaryClass = "hep-th",
    reportNumber = "MPP-2012-102, LMU-ASC-38-12, MPP--2012--102",
    doi = "10.1002/prop.201300002",
    journal = "Fortsch. Phys.",
    volume = "61",
    pages = "768--778",
    year = "2013"
}

@article{Li:2021utg,
    author = "Li, Yixuan",
    title = "{An Alliance in the Tripartite Conflict over Moduli Space}",
    eprint = "2112.03281",
    archivePrefix = "arXiv",
    primaryClass = "hep-th",
    month = "12",
    year = "2021"
}

@article{Blumenhagen:2019vgj,
    author = "Blumenhagen, Ralph and Brinkmann, Max and Makridou, Andriana",
    title = "{Quantum Log-Corrections to Swampland Conjectures}",
    eprint = "1910.10185",
    archivePrefix = "arXiv",
    primaryClass = "hep-th",
    reportNumber = "MPP-2019-204",
    doi = "10.1007/JHEP02(2020)064",
    journal = "JHEP",
    volume = "02",
    pages = "064",
    year = "2020"
}

@article{Ihl:2007ah,
    author = "Ihl, Matthias and Robbins, Daniel and Wrase, Timm",
    title = "{Toroidal orientifolds in IIA with general NS-NS fluxes}",
    eprint = "0705.3410",
    archivePrefix = "arXiv",
    primaryClass = "hep-th",
    reportNumber = "UTTG-02-07",
    doi = "10.1088/1126-6708/2007/08/043",
    journal = "JHEP",
    volume = "08",
    pages = "043",
    year = "2007"
}

@article{Bai:2021ewf,
    author = "Bai, Yang and Lu, Sida and Orlofsky, Nicholas",
    title = "{Searching for Magnetic Monopoles with the Earth\textquoteright{}s Magnetic Field}",
    eprint = "2103.06286",
    archivePrefix = "arXiv",
    primaryClass = "hep-ph",
    doi = "10.1103/PhysRevLett.127.101801",
    journal = "Phys. Rev. Lett.",
    volume = "127",
    number = "10",
    pages = "101801",
    year = "2021"
}

@article{Raju:2020smc,
    author = "Raju, Suvrat",
    title = "{Lessons from the information paradox}",
    eprint = "2012.05770",
    archivePrefix = "arXiv",
    primaryClass = "hep-th",
    doi = "10.1016/j.physrep.2021.10.001",
    journal = "Phys. Rept.",
    volume = "943",
    pages = "1--80",
    year = "2022"
}

@article{Calderon-Infante:2025pls,
    author = "Calder\'on-Infante, Jos\'e and Delgado, Matilda and Li, Yixuan and Lust, Dieter and Uranga, Angel M.",
    title = "{Classical Black Hole Probes of UV Scales}",
    eprint = "2502.03514",
    archivePrefix = "arXiv",
    primaryClass = "hep-th",
    reportNumber = "CERN-TH-2025-033, MPP-2025-10, LMU-ASC 04/25, IFT-UAM/CSIC-25-009",
    month = "2",
    year = "2025"
}

@article{Estes:2022buj,
    author = "Estes, John and Kavic, Michael and Liebling, Steven L. and Lippert, Matthew and Simonetti, John H.",
    title = "{Stability and observability of magnetic primordial black hole-neutron star collisions}",
    eprint = "2209.06060",
    archivePrefix = "arXiv",
    primaryClass = "astro-ph.HE",
    reportNumber = "OW-2022-2",
    doi = "10.1088/1475-7516/2023/06/017",
    journal = "JCAP",
    volume = "06",
    pages = "017",
    year = "2023"
}

@article{Ghosh:2020tdu,
    author = "Ghosh, Diptimoy and Thalapillil, Arun and Ullah, Farman",
    title = "{Astrophysical hints for magnetic black holes}",
    eprint = "2009.03363",
    archivePrefix = "arXiv",
    primaryClass = "hep-ph",
    doi = "10.1103/PhysRevD.103.023006",
    journal = "Phys. Rev. D",
    volume = "103",
    number = "2",
    pages = "023006",
    year = "2021"
}

@article{Bai:2020spd,
    author = "Bai, Yang and Berger, Joshua and Korwar, Mrunal and Orlofsky, Nicholas",
    title = "{Phenomenology of magnetic black holes with electroweak-symmetric coronas}",
    eprint = "2007.03703",
    archivePrefix = "arXiv",
    primaryClass = "hep-ph",
    reportNumber = "PITT-PACC-2005",
    doi = "10.1007/JHEP10(2020)210",
    journal = "JHEP",
    volume = "10",
    pages = "210",
    year = "2020"
}

@article{Cribiori:2023ffn,
    author = "Cribiori, Niccol\`o and Lust, Dieter and Montella, Carmine",
    title = "{Species entropy and thermodynamics}",
    eprint = "2305.10489",
    archivePrefix = "arXiv",
    primaryClass = "hep-th",
    reportNumber = "LMU-ASC 18/23, MPP-2023-97",
    doi = "10.1007/JHEP10(2023)059",
    journal = "JHEP",
    volume = "10",
    pages = "059",
    year = "2023"
}

@article{Diamond:2021scl,
    author = "Diamond, Melissa D. and Kaplan, David E.",
    title = "{Constraints on relic magnetic black holes}",
    eprint = "2103.01850",
    archivePrefix = "arXiv",
    primaryClass = "hep-ph",
    doi = "10.1007/JHEP03(2022)157",
    journal = "JHEP",
    volume = "03",
    pages = "157",
    year = "2022"
}

@article{Maldacena:2020skw,
    author = "Maldacena, Juan",
    title = "{Comments on magnetic black holes}",
    eprint = "2004.06084",
    archivePrefix = "arXiv",
    primaryClass = "hep-th",
    doi = "10.1007/JHEP04(2021)079",
    journal = "JHEP",
    volume = "04",
    pages = "079",
    year = "2021"
}

@article{Gervalle:2024yxj,
    author = "Gervalle, Romain and Volkov, Mikhail S.",
    title = "{Black Holes with Electroweak Hair}",
    eprint = "2406.14357",
    archivePrefix = "arXiv",
    primaryClass = "hep-th",
    doi = "10.1103/PhysRevLett.133.171402",
    journal = "Phys. Rev. Lett.",
    volume = "133",
    number = "17",
    pages = "171402",
    year = "2024"
}

@article{Herraez:2024kux,
    author = {Herr\'aez, Alvaro and L\"ust, Dieter and Masias, Joaquin and Scalisi, Marco},
    title = "{On the Origin of Species Thermodynamics and the Black Hole - Tower Correspondence}",
    eprint = "2406.17851",
    archivePrefix = "arXiv",
    primaryClass = "hep-th",
    reportNumber = "MPP-2024-123, LMU-ASC 08/24",
    month = "6",
    year = "2024"
}

@article{Basile:2024dqq,
    author = "Basile, Ivano and Cribiori, Niccol\`o and Lust, Dieter and Montella, Carmine",
    title = "{Minimal black holes and species thermodynamics}",
    eprint = "2401.06851",
    archivePrefix = "arXiv",
    primaryClass = "hep-th",
    reportNumber = "LMU-ASC 02/24, MPP-2024-6",
    doi = "10.1007/JHEP06(2024)127",
    journal = "JHEP",
    volume = "06",
    pages = "127",
    year = "2024"
}

@article{Horne:1992bi,
    author = "Horne, James H. and Horowitz, Gary T.",
    title = "{Black holes coupled to a massive dilaton}",
    eprint = "hep-th/9210012",
    archivePrefix = "arXiv",
    reportNumber = "UCSBTH-92-17",
    doi = "10.1016/0550-3213(93)90621-U",
    journal = "Nucl. Phys. B",
    volume = "399",
    pages = "169--196",
    year = "1993"
}

@article{Camara:2005dc,
    author = "Camara, Pablo G. and Font, A. and Ibanez, L. E.",
    title = "{Fluxes, moduli fixing and MSSM-like vacua in a simple IIA orientifold}",
    eprint = "hep-th/0506066",
    archivePrefix = "arXiv",
    reportNumber = "IFT-UAM-CSIC-05-28",
    doi = "10.1088/1126-6708/2005/09/013",
    journal = "JHEP",
    volume = "09",
    pages = "013",
    year = "2005"
}

@article{Lust:2004ig,
    author = "Lust, Dieter and Tsimpis, Dimitrios",
    title = "{Supersymmetric AdS(4) compactifications of IIA supergravity}",
    eprint = "hep-th/0412250",
    archivePrefix = "arXiv",
    reportNumber = "MPP-2004-173",
    doi = "10.1088/1126-6708/2005/02/027",
    journal = "JHEP",
    volume = "02",
    pages = "027",
    year = "2005"
}

@article{Becker:2024ayh,
    author = "Becker, Katrin and Brady, Nathan and Gra\~na, Mariana and Morros, Miguel and Sengupta, Anindya and You, Qi",
    title = "{Tadpole conjecture in non-geometric backgrounds}",
    eprint = "2407.16758",
    archivePrefix = "arXiv",
    primaryClass = "hep-th",
    doi = "10.1007/JHEP10(2024)021",
    journal = "JHEP",
    volume = "10",
    pages = "021",
    year = "2024"
}

@article{Bonnefoy:2019nzv,
    author = {Bonnefoy, Quentin and Ciambelli, Luca and L\"ust, Dieter and L\"ust, Severin},
    title = "{Infinite Black Hole Entropies at Infinite Distances and Tower of States}",
    eprint = "1912.07453",
    archivePrefix = "arXiv",
    primaryClass = "hep-th",
    reportNumber = "LMU-ASC 54/19, IPhT-T19/163, CPHT096.122019, MPP-2019-257, DESY 19-226, DESY-19-226, LMU\textendash{}ASC 54/19",
    doi = "10.1016/j.nuclphysb.2020.115112",
    journal = "Nucl. Phys. B",
    volume = "958",
    pages = "115112",
    year = "2020"
}

@article{DeWolfe:2005uu,
    author = "DeWolfe, Oliver and Giryavets, Alexander and Kachru, Shamit and Taylor, Washington",
    title = "{Type IIA moduli stabilization}",
    eprint = "hep-th/0505160",
    archivePrefix = "arXiv",
    reportNumber = "MIT-CTP-3640, PUPT-2161, SU-ITP-05-16, SLAC-PUB-11153",
    doi = "10.1088/1126-6708/2005/07/066",
    journal = "JHEP",
    volume = "07",
    pages = "066",
    year = "2005"
}

@article{Derendinger:2004jn,
    author = "Derendinger, Jean-Pierre and Kounnas, Costas and Petropoulos, P. Marios and Zwirner, Fabio",
    title = "{Superpotentials in IIA compactifications with general fluxes}",
    eprint = "hep-th/0411276",
    archivePrefix = "arXiv",
    reportNumber = "NEIP-04-08, LPTENS-0440, CPTH-RR053-0904, CERN-PH-TH-2004-228, ROMA-1395-04",
    doi = "10.1016/j.nuclphysb.2005.02.038",
    journal = "Nucl. Phys. B",
    volume = "715",
    pages = "211--233",
    year = "2005"
}

@article{Ihl:2006pp,
    author = "Ihl, Matthias and Wrase, Timm",
    title = "{Towards a Realistic Type IIA T**6/Z(4) Orientifold Model with Background Fluxes. Part 1. Moduli Stabilization}",
    eprint = "hep-th/0604087",
    archivePrefix = "arXiv",
    reportNumber = "UTTG-05-06",
    doi = "10.1088/1126-6708/2006/07/027",
    journal = "JHEP",
    volume = "07",
    pages = "027",
    year = "2006"
}

@article{Grimm:2004ua,
    author = "Grimm, Thomas W. and Louis, Jan",
    title = "{The Effective action of type IIA Calabi-Yau orientifolds}",
    eprint = "hep-th/0412277",
    archivePrefix = "arXiv",
    doi = "10.1016/j.nuclphysb.2005.04.007",
    journal = "Nucl. Phys. B",
    volume = "718",
    pages = "153--202",
    year = "2005"
}

@article{Delgado:2022dkz,
    author = "Delgado, Matilda and Montero, Miguel and Vafa, Cumrun",
    title = "{Black holes as probes of moduli space geometry}",
    eprint = "2212.08676",
    archivePrefix = "arXiv",
    primaryClass = "hep-th",
    reportNumber = "IFT-UAM/CSIC-22-151",
    doi = "10.1007/JHEP04(2023)045",
    journal = "JHEP",
    volume = "04",
    pages = "045",
    year = "2023"
}

@article{Long:2021lon,
    author = "Long, Cody and Sheshmani, Artan and Vafa, Cumrun and Yau, Shing-Tung",
    title = "{Non-Holomorphic Cycles and Non-BPS Black Branes}",
    eprint = "2104.06420",
    archivePrefix = "arXiv",
    primaryClass = "hep-th",
    reportNumber = "CIMP-D-21-00618R0",
    doi = "10.1007/s00220-022-04587-4",
    journal = "Commun. Math. Phys.",
    volume = "399",
    number = "3",
    pages = "1991--2043",
    year = "2023"
}

@article{Maldacena:2001xj,
    author = "Maldacena, Juan Martin and Moore, Gregory W. and Seiberg, Nathan",
    title = "{D-brane instantons and K theory charges}",
    eprint = "hep-th/0108100",
    archivePrefix = "arXiv",
    reportNumber = "RUNHETC-2001-24",
    doi = "10.1088/1126-6708/2001/11/062",
    journal = "JHEP",
    volume = "11",
    pages = "062",
    year = "2001"
}

@article{Witten:1998xy,
    author = "Witten, Edward",
    title = "{Baryons and branes in anti-de Sitter space}",
    eprint = "hep-th/9805112",
    archivePrefix = "arXiv",
    reportNumber = "IASSNS-HEP-98-42",
    doi = "10.1088/1126-6708/1998/07/006",
    journal = "JHEP",
    volume = "07",
    pages = "006",
    year = "1998"
}

@article{EnriquezRojo:2020hzi,
    author = "Enr\'\i{}quez Rojo, Mart\'\i{}n and Plauschinn, Erik",
    title = "{Swampland conjectures for type IIB orientifolds with closed-string U(1)s}",
    eprint = "2002.04050",
    archivePrefix = "arXiv",
    primaryClass = "hep-th",
    doi = "10.1007/JHEP07(2020)026",
    journal = "JHEP",
    volume = "07",
    pages = "026",
    year = "2020"
}

@article{Grana:2005jc,
    author = "Grana, Mariana",
    title = "{Flux compactifications in string theory: A Comprehensive review}",
    eprint = "hep-th/0509003",
    archivePrefix = "arXiv",
    reportNumber = "LPTENS-05-26, CPHT-RR-049-0805",
    doi = "10.1016/j.physrep.2005.10.008",
    journal = "Phys. Rept.",
    volume = "423",
    pages = "91--158",
    year = "2006"
}

@article{Grimm:2005fa,
    author = "Grimm, Thomas W.",
    title = "{The Effective action of type II Calabi-Yau orientifolds}",
    eprint = "hep-th/0507153",
    archivePrefix = "arXiv",
    reportNumber = "DESY-THESIS-2005-020, ZMP-HH-05-14",
    doi = "10.1002/prop.200510253",
    journal = "Fortsch. Phys.",
    volume = "53",
    pages = "1179--1271",
    year = "2005"
}

@article{Ooguri:2006in,
    author = "Ooguri, Hirosi and Vafa, Cumrun",
    title = "{On the Geometry of the String Landscape and the Swampland}",
    eprint = "hep-th/0605264",
    archivePrefix = "arXiv",
    reportNumber = "CALT-68-2600, HUTP-06-A017",
    doi = "10.1016/j.nuclphysb.2006.10.033",
    journal = "Nucl. Phys. B",
    volume = "766",
    pages = "21--33",
    year = "2007"
}

@article{Angius:2023xtu,
    author = "Angius, Roberta and Huertas, Jesus and Uranga, Angel M.",
    title = "{Small Black Hole Explosions}",
    eprint = "2303.15903",
    archivePrefix = "arXiv",
    primaryClass = "hep-th",
    reportNumber = "IFT-UAM/CSIC-23-31",
    month = "3",
    year = "2023"
}

@article{Danielsson:2006jg,
    author = "Danielsson, Ulf H. and Johansson, Niklas and Larfors, Magdalena",
    title = "{Stability of flux vacua in the presence of charged black holes}",
    eprint = "hep-th/0605106",
    archivePrefix = "arXiv",
    reportNumber = "UUITP-07-06",
    doi = "10.1088/1126-6708/2006/09/069",
    journal = "JHEP",
    volume = "09",
    pages = "069",
    year = "2006"
}

@article{Green:2006nv,
    author = "Green, Daniel R. and Silverstein, Eva and Starr, David",
    title = "{Attractor explosions and catalyzed vacuum decay}",
    eprint = "hep-th/0605047",
    archivePrefix = "arXiv",
    reportNumber = "SLAC-PUB-11846, SU-ITP-06-13",
    doi = "10.1103/PhysRevD.74.024004",
    journal = "Phys. Rev. D",
    volume = "74",
    pages = "024004",
    year = "2006"
}

@article{Cribiori:2022cho,
    author = "Cribiori, Niccol\`o and Dierigl, Markus and Gnecchi, Alessandra and Lust, Dieter and Scalisi, Marco",
    title = "{Large and small non-extremal black holes, thermodynamic dualities, and the Swampland}",
    eprint = "2202.04657",
    archivePrefix = "arXiv",
    primaryClass = "hep-th",
    reportNumber = "LMU-ASC 05/22, MPP-2022-20",
    doi = "10.1007/JHEP10(2022)093",
    journal = "JHEP",
    volume = "10",
    pages = "093",
    year = "2022"
}

@article{Bacchini:2021fig,
    author = "Bacchini, Fabio and Mayerson, Daniel R. and Ripperda, Bart and Davelaar, Jordy and Olivares, H\'ector and Hertog, Thomas and Vercnocke, Bert",
    title = "{Fuzzball Shadows: Emergent Horizons from Microstructure}",
    eprint = "2103.12075",
    archivePrefix = "arXiv",
    primaryClass = "hep-th",
    doi = "10.1103/PhysRevLett.127.171601",
    journal = "Phys. Rev. Lett.",
    volume = "127",
    number = "17",
    pages = "171601",
    year = "2021"
}

@article{Apers:2022vfp,
    author = "Apers, Fien",
    title = "{Aspects of AdS flux vacua with integer conformal dimensions}",
    eprint = "2211.04187",
    archivePrefix = "arXiv",
    primaryClass = "hep-th",
    doi = "10.1007/JHEP05(2023)040",
    journal = "JHEP",
    volume = "05",
    pages = "040",
    year = "2023"
}

\end{document}